\documentclass{pazhb_eng}
\usepackage{graphicx}
\usepackage{natbib}
\usepackage{color}

\usepackage{lscape}
\usepackage{amssymb}
\usepackage{amsmath}

\def\arcsec{$^{\prime\prime\,}$}
\def\arcmin{$^{\prime\,}$}

\def\flux{erg s$^{-1}$ cm$^{-2}$}
\def\ergs{erg s$^{-1}$}

\begin{document}

\journalinfo{2015}{41}{5}{179}[195]

\title{Determining the Nature of Faint X-Ray Sources from the ASCA Galactic
Center Survey}

\author{A.\,A.\,Lutovinov\email{aal@iki.rssi.ru}\address{1}, M.\,G.\,Revnivtsev\address{1}, D.\,I.\,Karasev\address{1},
V.\,V.\,Shimansky\address{2}, R.\,A.\,Burenin\address{1}, I.\,F.\,Bikmaev\address{2,3},	V.\,S.\,Vorobyev\address{1},
S.\,S.\,Tsygankov\address{4}, M.\,N.\,Pavlinsky\address{1}\\
    \bigskip
  {\it (1) Space Research Institute, Russian Academy of Sciences, Moscow, Russia}\\
  {\it (2) Kazan (Volga Region) Federal University, Kazan, Russia}\\
  {\it (3) Academy of Sciences of Tatarstan, Kazan, Russia}\\
  {\it (4) Tuorla Observatory, University of Turku, Finland}\\
}

\shortauthor{Lutovinov et al.}

\shorttitle{Determining the Nature of Faint X-Ray Sources}

\submitted{9 December 2014}

\begin{abstract}

We present the results of the identification of six objects from the ASCA
Galactic center and Galactic plane surveys: AX\,J173548-3207, AX\,J173628-3141,
AX\,J1739.5-2910, AX\,J1740.4-2856, AX\,J1740.5-2937, AX\,J1743.9-2846. Chandra, XMM-Newton, and {\it XRT}/Swift
X-ray data have been used to improve the positions of the optical counterparts to these sources. Thereafter,
we have carried out a series of spectroscopic observations of the established optical counterparts at the
RTT-150 telescope. Analysis of X-ray and optical spectra as well as photometric measurements in a wide
wavelength range based on optical and infrared catalogs has allowed the nature of the program sources to
be determined.
Two X-ray objects have been detected in the error circle of  AX\,J173628-3141:
one is a coronally active G star and the other may be a symbiotic star, a red giant with an accreting white dwarf.
Three sources (AX\,J1739.5-2910, AX\,J1740.5-2937, AX\,J1743.9-2846) have turned out to be active G–K
stars, presumably RS CVn objects, one (AX\,J1740.4-2856) is an M dwarf, and another one (AX\,J173548-
3207) may be a low-mass X-ray binary in its low state. The distances and corresponding luminosities of the
sources in the soft X-ray band have been estimated; analysis of deep INTEGRAL Galactic
center observations has not revealed a statistically significant flux at energies $>$20 keV from any of them.

\englishkeywords{X-ray sources, active stars, Galaxy, ASCA observatory}

\end{abstract}

\section{Introduction}

The stellar population of galaxies consists of objects of various classes each of which can be an
X-ray source through physical processes of different nature. In particular, in binary systems with
neutron stars, black holes, and white dwarfs, X-ray emission is generated during the accretion of matter
onto compact objects; stars of late spectral types have active coronas in which matter heated to millions
and tens of millions degrees is an origin of X-ray emission; young stars can heat up matter in strong
wind flows emitted by them; etc. Analysis of the physical processes in these objects often has to be
made by studying a whole population of sources of a certain class, which makes it necessary to perform
sky surveys and to identify detected sources.

In view of different design features of the instruments
capable of performing all-sky surveys and despite their continuously improving characteristics,
so far even the most sensitive RXTE (Revnivtsev et al. 2004), INTEGRAL (Krivonos et al. 2007),
and Swift (Baumgartner et al. 2013) all-sky surveys have a sensitivity $10^{-11}$ \flux. The
ROSAT sky survey performed at a sensitivity of $10^{-12}$ \flux\ (Voges et al. 1999) constitutes
an exception. However, this survey was performed at energies below 2 keV, an energy range strongly affected by
interstellar absorption, which significantly limited its capabilities to study objects in the Galactic disk.

Observations with more sensitive instruments within the framework of special survey programs (see,
e.g., Wang et al. 2002; Hands et al. 2004; Ebisawa et al. 2005; van den Berg et al. 2009; Revnivtsev
et al. 2009) or programs to identify and observe X-ray sources (Kim et al. 2004; Saxton et al. 2008;
Evans et al. 2010, 2014; Warwick 2014) also make it possible to obtain surveys of some Galactic regions.
However, since the field of view of focal telescopes is small, they all cover only small areas in the sky. So
far the ASCA surveys (Sugizaki et al. 2001; Sakano et al. 2002) have been among the most extensive (in
Galactic center and Galactic plane coverage) surveys performed by X-ray telescopes with focusing optics
at energies above 2 keV.

However, these surveys have also a significant shortcoming: a substantial fraction of the objects
detected in these surveys are of unknown nature in view of the insufficient ASCA angular resolution,
which makes it difficult to use them to study various populations of objects in our Galaxy. In addition,
 the analysis of ASCA data, the search for sources, and their detection have been
significantly hampered by the proper allowance for the contribution from the Galactic ridge X-ray emission,
allowance for the influence of bright objects, etc. The latest-generation Chandra and XMM-Newton X-ray
observatories have carried out systematic observations of the sky fields in which sources were
detected in the ASCA surveys for a long time. This allows one to improve considerably the astrometric
positions of X-ray sources and then to carry out optical or infrared observations to determine their nature
(Degenaar et al. 2012; Anderson et al. 2014).

In a harder X-ray band ($>17$ KeV), the most sensitive Galactic center and Galactic plane survey is
the INTEGRAL one (Krivonos et al. 2012). This survey has a high identification completeness of detected
sources ($\sim92$\%) and a deep and fairly uniform (in exposure) coverage of the inner part of the
Galactic plane, which allows statistical and population studies of X-ray sources of various classes to
be performed (see, e.g., Revnivtsev et al. 2008 and Lutovinov et al. 2013, for low-mass and high-mass
X-ray binaries, respectively). The Galactic center region is of particular interest for investigating the
spatial distribution of high-mass X-ray binaries. This is because, according to the theory of stellar evolution,
such objects representing the young population of the Galaxy must concentrate toward the regions of
enhanced star formation, i.e., toward the spiral arms. Currently available observational data confirm this
viewpoint (see, e.g., Grimm et al. 2002; Lutovinov et al. 2005; Bodaghee et al. 2012; and references
therein). At the same time, it should be noted that high-mass X-ray binaries can, in principle, also exist
in galactic bulges (a bulge is a thickening observed at the centers of spiral galaxies), in regions where
star formation processes take place (for example, in molecular clouds), but so far there are no unambiguous
observational confirmations of this. Within the framework of our program for search high-mass X-ray
binaries in the Galactic bulge, we are performing a systematic analysis of all the unidentified Xray
sources detected in the ASCA and INTEGRAL Galactic center surveys to establish their nature and
to determine distances to them. We have selected a total of 17 sources from the ASCA survey and
7 objects from the INTEGRAL survey for this study.

In this paper, we present the results of the first
stage of this program within the framework of which we have identified six faint objects from the ASCA
survey with the help of spectroscopic observations of candidates at the Russian-Turkish 1.5-m telescope
(RTT-150) using the {\it TFOSC} medium- and low-resolution spectrometer.

\section{X-ray observations}

Figure 1 shows the map of the Galactic center
region with coordinates $-4^\circ\le l \le 4^\circ$, $-1.3^\circ\le b \le 1.3^\circ$,
obtained by summing all the available INTEGRAL data over the period of observations 2003–
2014 in the 17–60 keV energy band (the total exposure time is more than 40 Ms). Darker regions
correspond to the positions of brighter objects in hard X-rays.

The positions of the sources detected in the ASCA surveys are indicated by circles of different colors:
the green color indicates the sources whose nature is known (their widely used names rather than the
names from the surveys themselves are given); the red color marks the objects whose nature has not
yet been established. A total of 42 sources detected by ASCA in this sky field are shown on the map,
for 25 of which the type and class have already been determined, while the nature of 17 more sources is
unknown. Six of these 17 objects, which are the subject of our studies in this paper, are additionally
marked by the red squares. Note that all these sources are in the first quadrant and are distinguished by the
fact that the optical counterparts are optically bright stars. The latter has allowed us to
perform their optical spectroscopy and, in the long run, to determine their nature.

\begin{landscape}
\begin{figure*}
\hspace{-6.5cm}\includegraphics[width=1.35\textwidth,bb=36 308 575 486,clip]{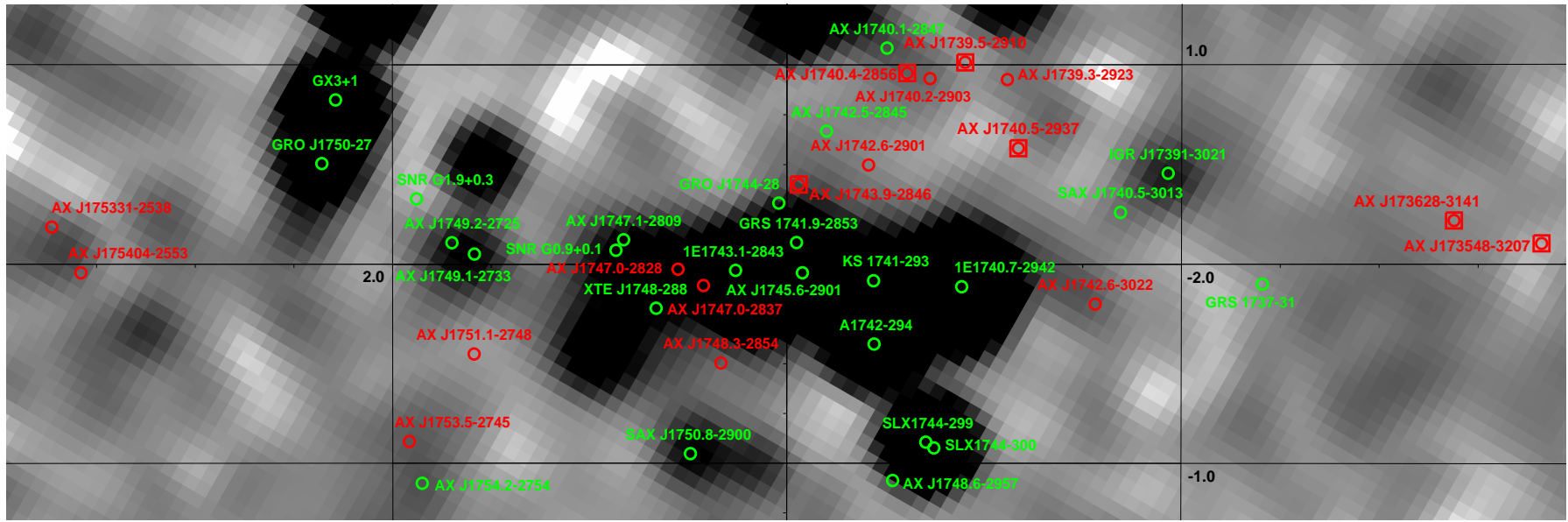}
\hspace{-6.5cm}\caption{INTEGRAL map of the Galactic center region with coordinates  $-4^\circ\le l \le 4^\circ$,
$-1.3^\circ\le b \le 1.3^\circ$, in the 17–60 keV energy band. The red circles indicate positions
of the sources of unknown nature from the ASCA surveys (Sugizaki et al. 2001; Sakano et al. 2002).
The objects being investigated in this paper are additionally marked by the red squares. The green
color indicates the sources of known nature from the same ASCA surveys. For the reader’s convenience,
the widely used names are given for them.\label{axs_int_image}}\hspace{6.5cm}
\end{figure*}
\end{landscape}

It can be seen from Fig. \ref{axs_int_image} that the signal from some of the objects detected by ASCA is also
registered in hard X-rays, with this being also true for the sources of unknown nature. However, none of
the objects investigated in this paper was detected at a statistically significant level by INTEGRAL.
The ($3\sigma$) upper limits on the 20–40 keV fluxes from 

\begin{landscape}
\begin{table*}
\footnotesize
\caption{List of sources, their fluxes, spectral parameters, optical counterparts, distances, and luminosities }
~\hspace{-75mm}~ \begin{tabular}{l|l|c|c|c|c|c|c|c|c|c|c}
\hline[0.5mm]
Source & 2MASS  & ObsID  & \multicolumn{3}{c|}{$F_x, 10^{-13}$ \flux} & \multicolumn{2}{c|}{Spectral parameters} & $m_j$ & Type & Distance & $L_{X}$ \\[0.5mm]
\cline{4-8}
 & & ObsID    &  0.5-2 & 2-10 & 0.7-10 & $\Gamma$    & $kT$, KeV    & & & kpc & \ergs\ \\[0.5mm]
 & &          &   KeV  &   KeV&    KeV & $N_{\rm H}$ & $N_{\rm H}$  & & &     &        \\[0.5mm]
\hline
AX\,J173548-3207 & J17354627-3207099 & Chandra & $0.64^{+0.13}_{-0.24}$ & $1.4^{+0.3}_{-1.0}$  & $2.0^{+0.2}_{-1.6}$  & $2.0\pm0.5$ & & 14.19 & qLMXB? & $<2$ & $<10^{32}$\\
                 &                   &   8148  &                        &                      &                      & $0.3\pm0.2$ & &       &  & & \\[1.0mm]
AX\,J173628-3141 & J17362729-3141234 & Chandra & $0.12^{+0.01}_{-0.02}$ & $<0.002$ &$0.10^{+0.01}_{-0.04}$& & $0.50\pm0.14$ & 11.03& G0III & 1.4 & $\sim3\times 10^{30}$\\
                 &                   &   8158  &                        &                      &                  &              & &       &  & &  \\
                 & VVV               &         &                        & $2.5^{+1.0}_{-2.0}$  & $2.5^{+1.1}_{-2.0}$  & 2.0(fix)    & &  $18.46$& & &      \\
                 &                   &         &                        &                      &                  & $>20$  & &        &   & &   \\[1.0mm]
AX\,J1739.5-2910 & J17393122-2909532 & Chandra & $4.14^{+0.38}_{-0.41}$ &$3.59^{+0.22}_{-0.37}$&$7.54^{+0.66}_{-1.31}$& $2.83\pm0.17$& & 7.21& G-KIII & 0.4 & $\sim1.2\times10^{31}$ \\
                 &                   &  14887  &                        &                      &                  & $0.25\pm0.08$& &       &   & &    \\
                 &                   & Chandra & $6.7^{+1.1}_{-2.8}$   & $4.06^{+0.37}_{-0.73}$& $10.4^{+1.0}_{-2.7}$ & $3.23\pm0.0.35$& &      &  & &    \\
                 &                   &  8679   &                       &                       &                  & $0.29\pm0.12$ & &      &   & &    \\
                 &                   &XMM-Newton& $6.2^{+3.0}_{-1.5}$  & $4.69^{+0.47}_{-0.79}$& $10.5^{+0.9}_{-1.8}$ & $2.78\pm0.21$ & &      &  & &     \\
                 &                   &0304220101&                      &                       &                  & $0.14\pm0.06$ & &      &    & &   \\
                 &                   &XMM-Newton& $5.2^{+0.6}_{-0.8}$  & $4.25^{+0.43}_{-0.61}$& $9.0^{+0.9}_{-1.1}$  & $2.66\pm0.19$ & &      &   & &    \\
                 &                   &0304220301&                      &                       &                  & $0.12\pm0.05$ & &      &   & &    \\[1.0mm]
AX\,J1740.4-2856 & J17402384-2856527 &XMM-Newton&$2.94^{+0.28}_{-0.50}$& $0.11^{+0.03}_{-0.04}$&$2.16^{+0.32}_{-0.27}$&  & $0.86\pm0.08$ & 9.88 & M0-2 & $\sim0.056$& $\sim 10^{29}$\\
                 &                   &0511010701&                      &                       &                  &  &               &      &   & &    \\
                 &                   &XMM-Newton& $3.2^{+1.6}_{-2.4}$  & $0.15^{+0.02}_{-0.10}$& $2.3^{+0.1}_{-1.7}$  &  & $0.89\pm0.35$ &      &  & &  \\
                 &                   &030422060 &                      &                       &                  &  &               &      &  & &    \\[1.0mm]
AX\,J1740.5-2937 & J17403458-2937438 & {\em XRT}/Swift & $6.1^{+1.2}_{-2.9}$ & $0.26^{+0.25}_{-0.10}$ & $5.7^{+1.8}_{-2.4}$ & & $0.75\pm0.13$ & 9.31&G0-5IV& $\sim0.67$& $\sim2-5\times10^{31}$\\
                 &                   &00043602001&                        &                     &                 & & $0.3\pm0.2$   &       &  & &   \\[1.0mm]
AX\,J1743.9-2846 & J17435129-2846380 &XMM-Newton& $3.0^{+0.6}_{-0.6}$  & $3.65^{+0.87}_{-0.54}$ & $6.6^{+0.5}_{-1.1}$ & $2.38\pm0.22$ & & 8.54 & G5III& 1.1& $\sim3-9\times10^{31}$\\
                 &                   &0112971901&                      &                        &                  & $0.14\pm0.06$ & &      &    & 0.5 & $\sim0.7-2\times10^{31}$  \\
                 &                   &XMM-Newton&$1.51^{+0.22}_{-1.01}$& $0.9^{+0.3}_{-0.6}$    & $2.4^{+0.3}_{-1.6}$ &   $4.0\pm1.0$ & &      &   & &    \\
                 &                   &0112971201&                      &                        &                  & $0.77\pm0.38$ & &      &   & &    \\
                 &                   & Chandra  &$1.46^{+0.23}_{-0.35}$& $1.93^{+0.30}_{-0.21}$ &$3.35^{+0.35}_{-1.21}$& $2.55\pm0.19$& &      &   & &    \\
                 &                   &  8209    &                      &                        &                 & $0.31\pm0.08$ & &      &    & &   \\[0.5mm]
\hline
\end{tabular}
\end{table*}
\end{landscape}

\noindent four sources, AX\,J173548-3207, AX\,J173628-
3141, AX\,J1739.5-2910, and AX\,J1740.5-2937, are $\simeq0.2$ mCrab. Formally, emission with fluxes of
$0.21\pm0.06$ and $0.32\pm0.06$ mCrab is registered from position boxes of two other sources, 
AX\,J1740.4-2856 and AX\,J1743.9-2846, respectively. However, given the insufficiently high angular resolution of
the {\it IBIS}/INTEGRAL telescope (12\arcmin, Winkler et al 2003), additional studies have established that these fluxes do not
belong to the objects under study: the measured flux for AX\,J1743.9-2846 is most likely attributable to
the contribution of emission from the nearby source GRS\,1741.9-2853, while the emission from the error
circle of AX\,J1740.4-2856 actually belongs to a different source from the ASCA survey, AX\,J1740.1-2847 
(Mereminsky et al., in preparation).

The positional accuracies for the sources in the ASCA Galactic center and Galactic plane surveys
are 50\arcsec\ and 1\arcmin, respectively (Sugizaki et al. 2001; Sakano et al. 2002), which is insufficient to identify
the detected objects with optical and infrared sources and to determine their nature. Nevertheless, owing
to the data obtained in the special Chandra, XMM-Newton, and {\it XRT}/Swift Galactic center surveys,
we have been able to identify the sources detected by ASCA with objects in these surveys. It should
be noted that we took the astrometric positions of the corresponding objects from the Chandra/XMM-Newton/
XRT surveys (Evans et al. 2010, Degenaar et al. 2012, Evans et al. 2014), and the
3XMM-DR4 catalogue\footnote{http://xmmssc-www.star.le.ac.uk/Catalogue/xcat\_ public\_\-3XMM-DR4.html}. 
This allowed us to measure their X-ray spectra and fluxes, to compare  results obtained with the ASCA 
results, and to perform spectroscopic observations of the probable
optical counterparts to these systems. The table gives the names of the corresponding objects and
their J magnitudes from the 2MASS catalog and, for the hard counterpart to AX\,J173628-3141, from the
VVV/ESO (the VISTA telescope) survey.

\begin{figure*}
\vbox{
\hbox{
\includegraphics[width=0.48\textwidth,bb=24 298 565 692,clip]{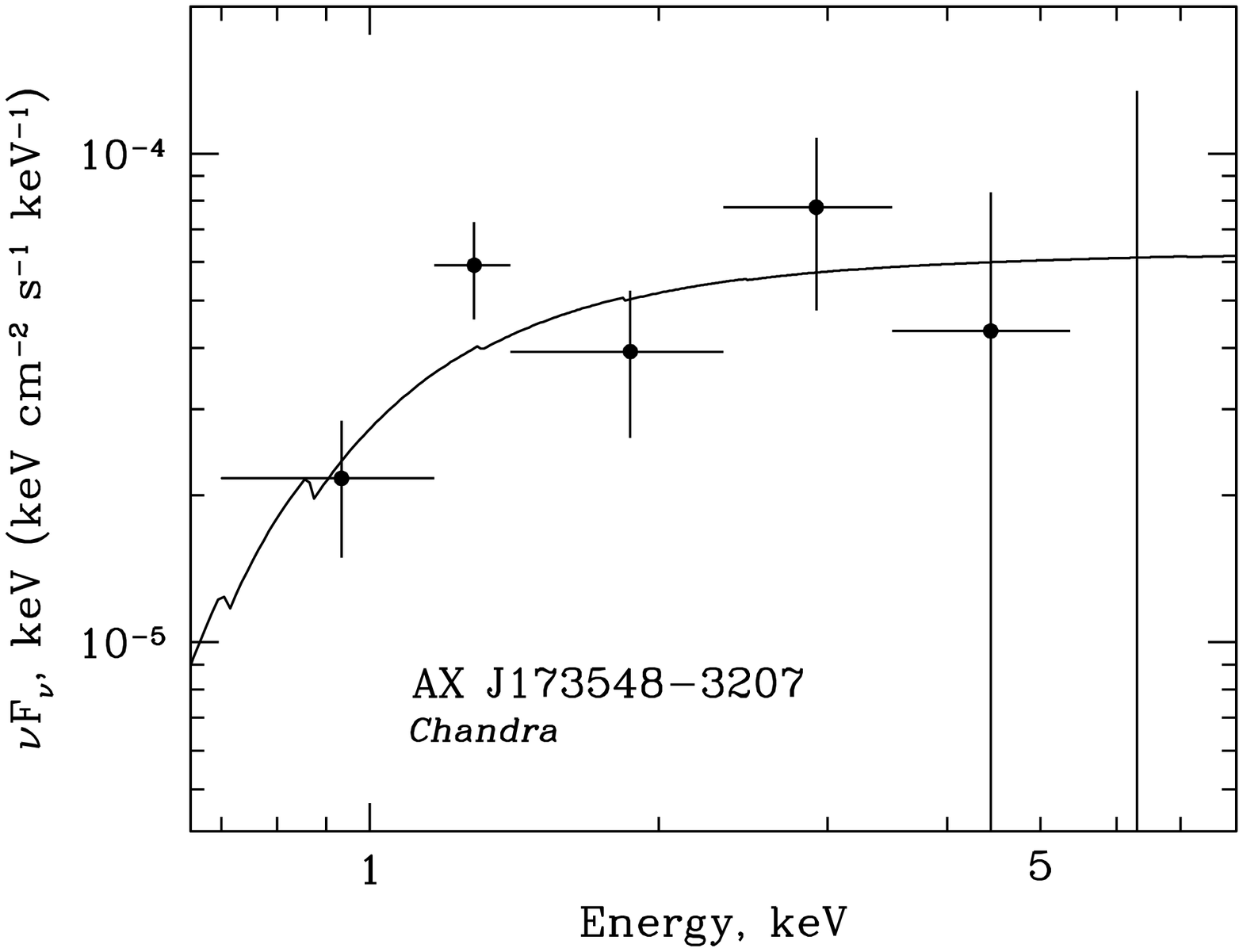}
\hspace{2mm}\includegraphics[width=0.48\textwidth,bb=24 298 565 692,clip]{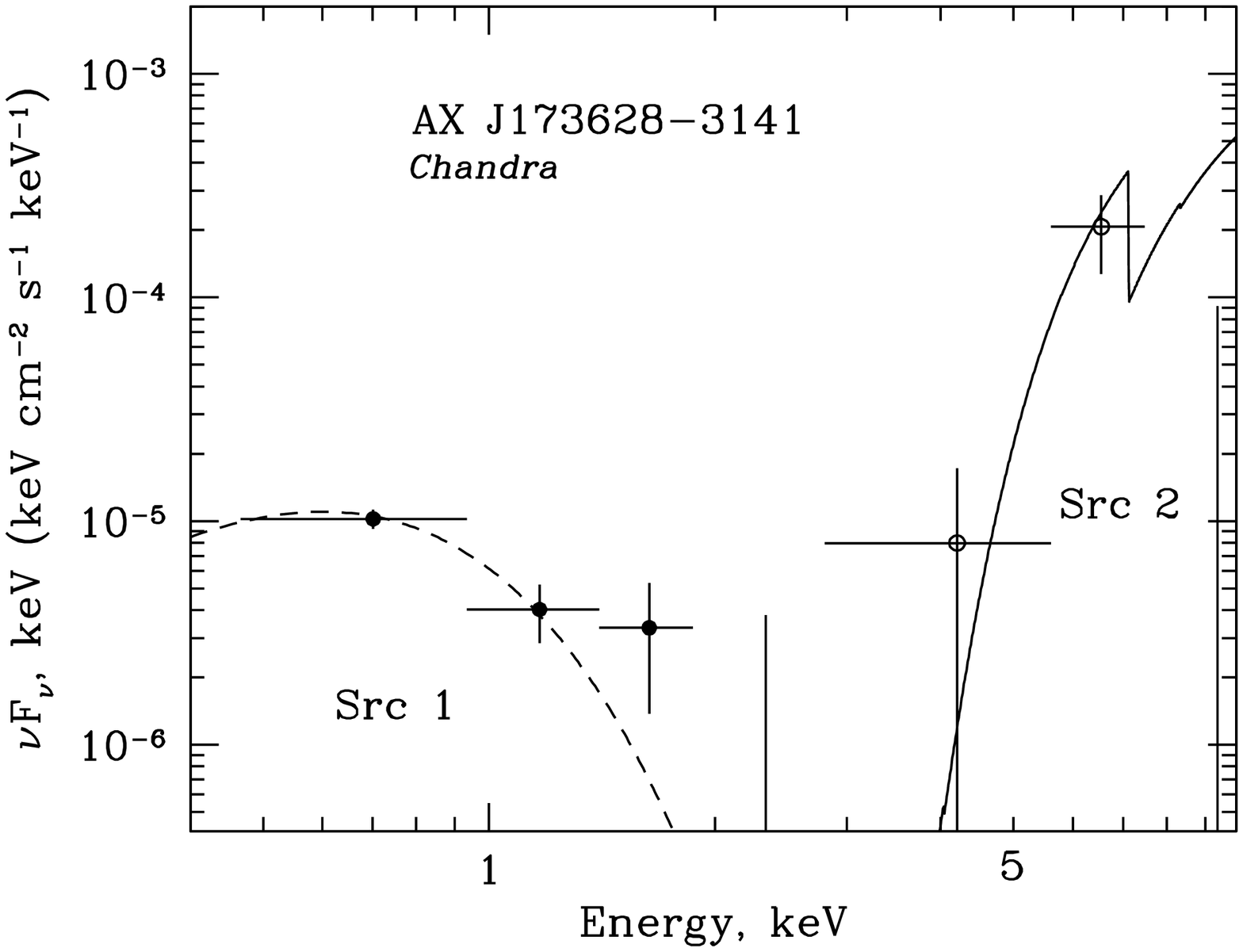}
}
\hbox{
\includegraphics[width=0.48\textwidth,bb=24 298 565 692,clip]{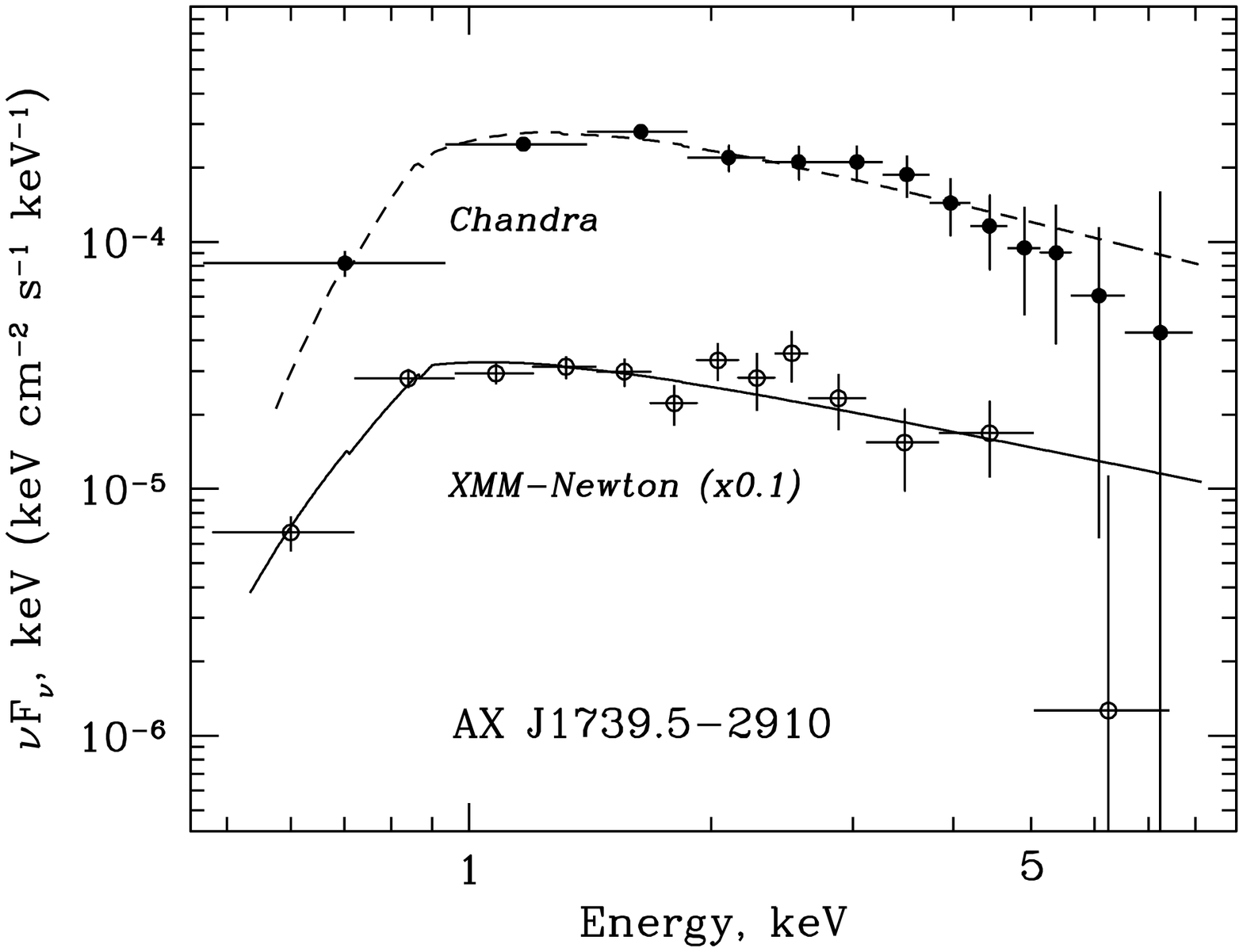}
\hspace{2mm}\includegraphics[width=0.48\textwidth,bb=24 298 565 692,clip]{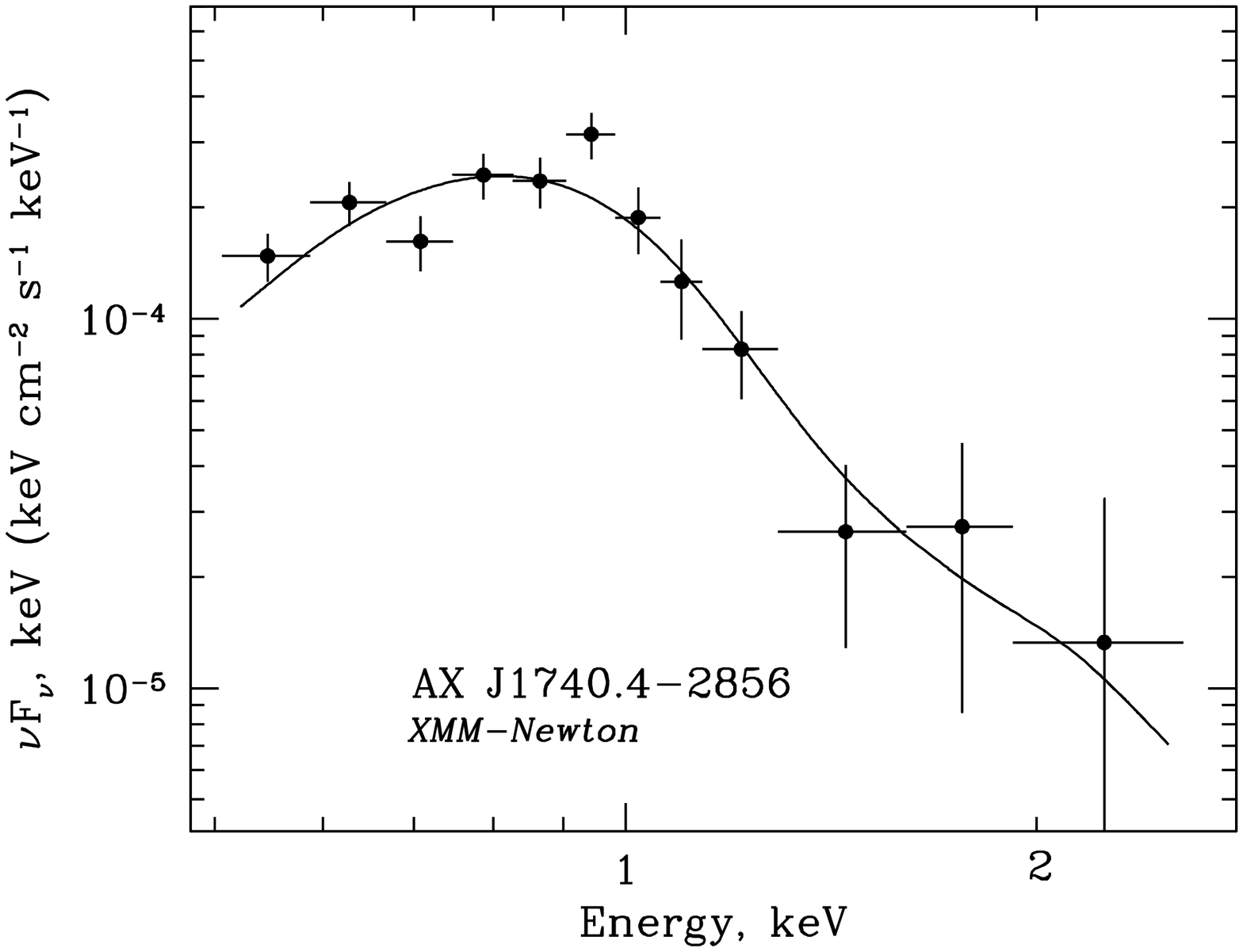}
}
\hbox{
\includegraphics[width=0.48\textwidth,bb=24 272 565 692,clip]{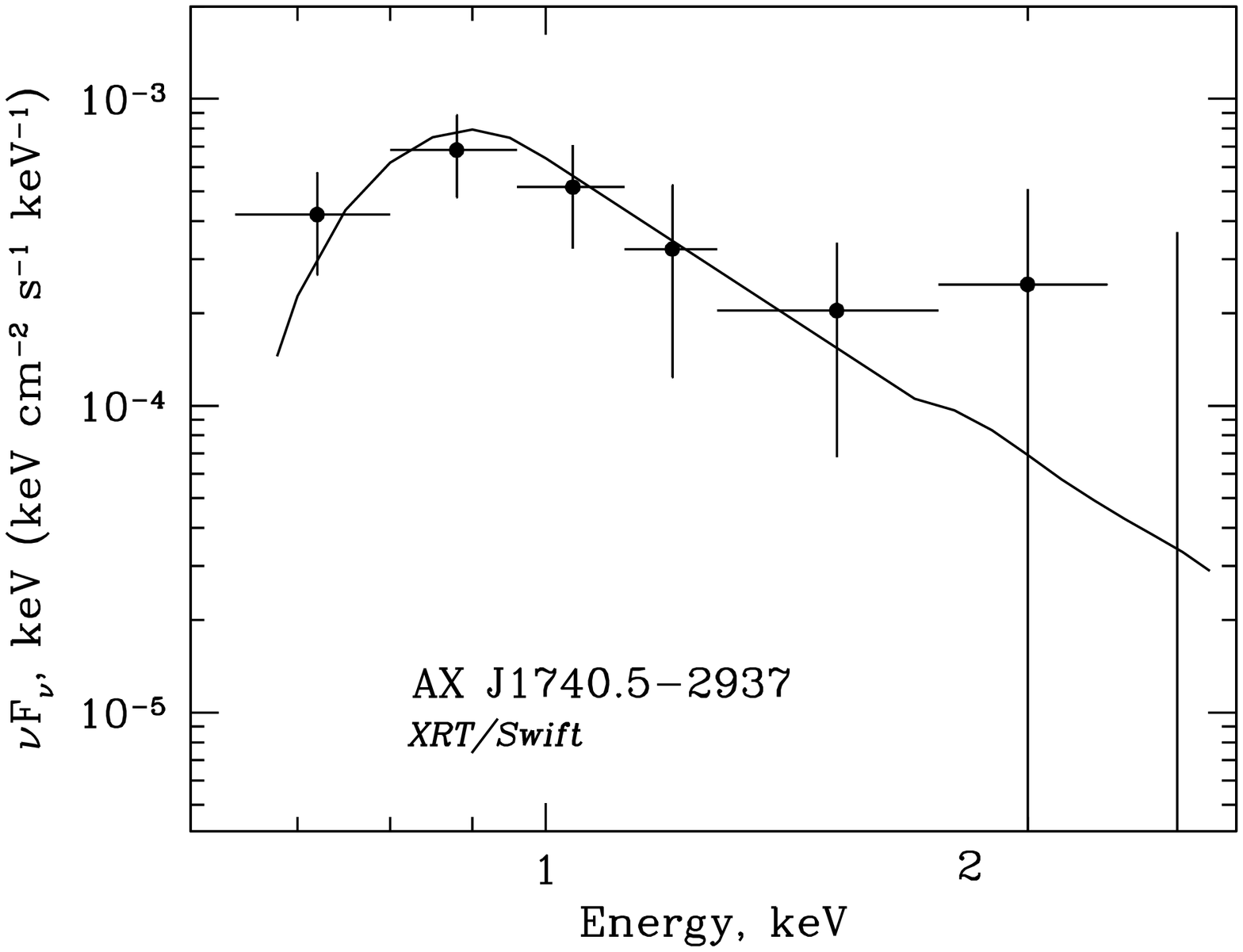}
\hspace{2mm}\includegraphics[width=0.48\textwidth,bb=24 272 565 692,clip]{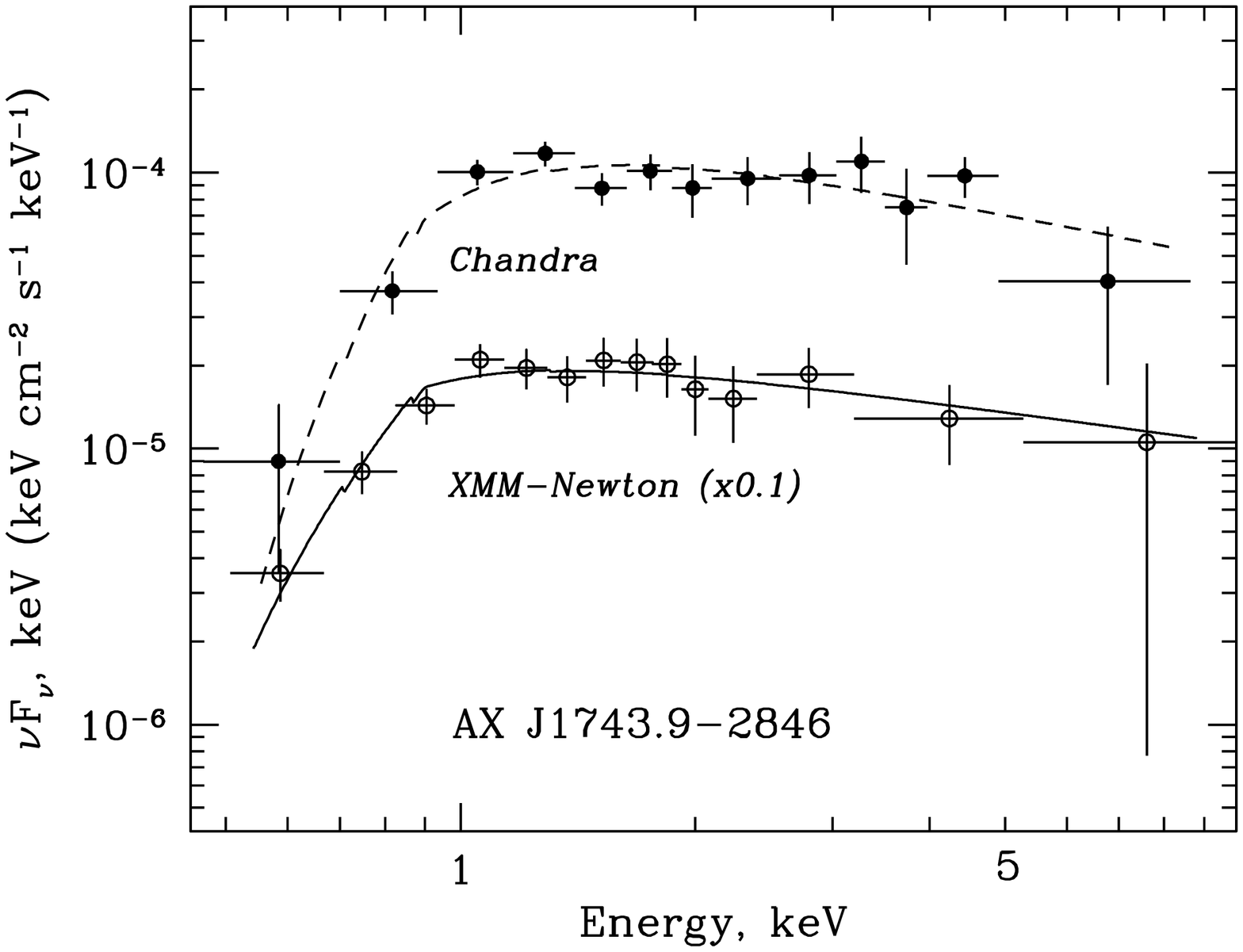}
}
}
\caption{Chandra, XMM-Newton, and {\em XRT}/Swift spectra of the studied sources in the 0.5–10 keV energy band. Solid
lines indicate the best fits to the spectra. The corresponding parameters are given in the table. }
\label{spectra}
\end{figure*}

In the next section, we present  results of our
optical observations and discuss the nature of each source separately, while here we will consider results
of spectral analysis in the X-ray wavelength range in general terms.
Figure \ref{spectra} shows the Chandra, XMM-Newton, and \emph{XRT}/Swift spectra of the six program sources in
the 0.5–10 keV energy band. The best fits to these spectra and the corresponding fluxes are presented in
the table. The fluxes are given in three energy bands: 0.5–2, 2–10, and 0.7–10 keV. The measurements
in the first two bands provide information about the ratio of the soft and hard X-ray fluxes, while the last
band allows the results of our measurements to be compared with the ASCA results obtained in the 0.7–
10 keV energy band  as well. The spectra of almost all sources can be satisfactorily described by a power law with
absorption at low energies; the corresponding values of $\Gamma$ and $N_{\rm H}$ are given in the table. For very soft
sources, in the spectra of which there are virtually no photons with energies $>$2 keV, we used the thermal
{\sc cevmkl} model of the multi-temperature plasma emission from the {\sc XSPEC} package; the corresponding
maximum temperature is given in the table. Note in advance that most of objects under study have
turned out to be active giants or subgiants of late spectral types (G–K), probably RS CVn binaries.
Therefore, to fit spectra of the brightest sources, AX\,J1739.5-2910 and AX\,J1743.9-2846, we used a
power law with a break in the form
$$
{A(E)}= \left\{ \begin{array}{rl}
K E^{-\Gamma_1},&\mbox{ if $E<E_{br}$}\\
K E_{br}^{\Gamma_2-\Gamma_1} E^{-\Gamma_2},&\mbox{if $E>E_{br}$}
\end{array} \right.
$$
\noindent where $\Gamma_1, \Gamma_2$ are the slopes of the spectrum before and after the break at energy $E_{br}$.
We fixed $\Gamma_1$ and $E_{br}$ at 0.5 and 0.9 keV, respectively, typical of them
(Sazonov et al. 2006), while the table gives $\Gamma_2$ and
interstellar absorption magnitudes.

To conclude this section, note that the construction of images from the {\em IBIS}/INTEGRAL data is
based on the balanced cross-correlation algorithm (for a description, see Krivonos et al. 2010; Churazov
et al. 2014); the Chandra and XMM-Newton data were analyzed with special-purpose software, {\sc CIAO 4.6} and
{\sc SAS v14.0.0}, respectively; the final data analysis and {\emph XRT} data processing were performed
with the{\sc FTOOLS/HEASOFT 6.14} package.

\begin{figure*}
\begin{center}
\includegraphics[width=0.95\textwidth,bb=35 307 575 486,clip]{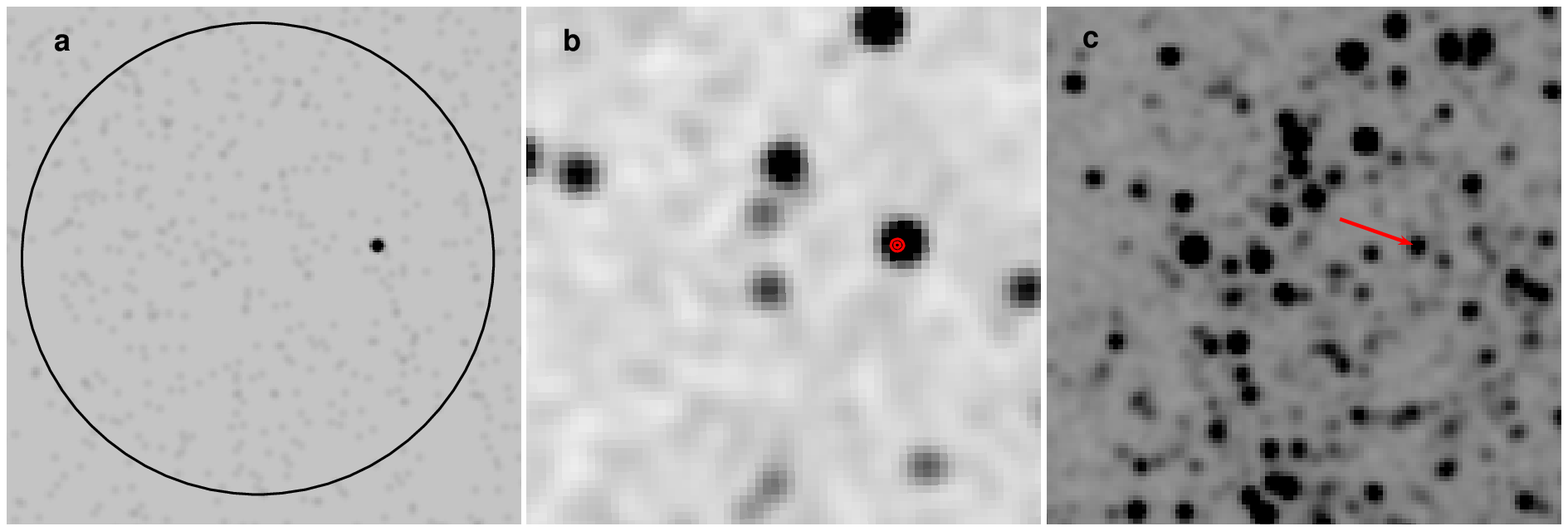}
\caption{Images of the sky field around AX\,J173548-3207. Panel (a) shows the Chandra image in the 0.5–8 keV X-ray band.
The circle indicates the ASCA positional accuracy of AX\,J173548-3207. The only X-ray source detected by Chandra coincides with
the bright optical (panel (b), the source's X-ray intensity distribution is indicated by the contours) and infrared (panel (c))
source 2MASS\,J17354627-3207099 (marked by the arrow).}
\label{axj173548_image}
\end{center}
\end{figure*}

\section{Results of optical identifications}

The optical spectra of the sources were taken with
the \emph{TFOSC} medium- and low-resolution spectrometer\footnote{http://hea.iki.rssi.ru/rtt150/ru/index.php?page=tfosc}
of RTT-150 using a grism that provided a resolution of about 12-15\AA\ and a high quantum efficiency
in a wide spectral range, 4000-9000\AA. Since all sources have declinations $-32^\circ<\textrm{Dec}<-28^\circ$
and are at very large zenith angles for RTT-150 (the telescope is located at latitude 36$^\circ$ 49$^\prime$ N,
and longitude 30$^\circ$ 20$^\prime$ E), the absolute accuracy
of our spectrophotometric measurements is low; therefore, we normalized the spectrum of an object
to its continuum for the subsequent analysis. The observations were carried out from May 16 to June 5,
2013, with an exposure time of 600–900 s per each source. 

\begin{figure}[htb]
\begin{center}
\includegraphics[width=\columnwidth,bb=50 185 570 490,clip]{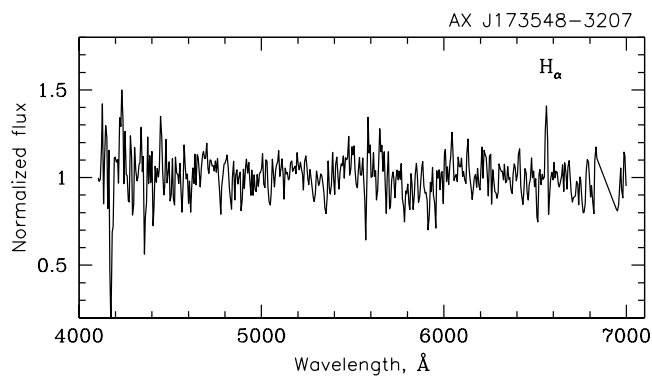}
\caption{RTT-150 optical spectrum (normalized to the continuum) of AX\,J173548-3207. }
\label{axj173548}
\end{center}
\end{figure}

\noindent The optical data were processed in a standard
way using the \emph{IRAF}\footnote{http://iraf.noao.edu} software package and our own software.

\subsection{AX\,J173548-3207}

AX\,J173548-3207 was observed by Chandra (Evans et al. 2010; Anderson et al. 2014) and fell several
times within the XRT/Swift field of view during the Swift Galactic plane survey (Evans et al. 2014).
The accurate astrometric position of the source allow it to be confidently associated with the object
2MASS\,J17354627-3207099 (Fig.\ref{axj173548_image}). Its optical spectrum at RTT-150 with an
exposure time of 900 s (Fig.\ref{axj173548}). The spectrum contains no distinct features, except for
the weak $H_\alpha$ (6563\AA) emission line. The presence of this emission line at $z=0$, suggests
that this source is a galactic one.

The optical spectrum suggests that source is most likely a low-mass X-ray binary system
(see, e.g., the spectrum of the low-mass X-ray binary EXO\,0748-676 in its low state; Bassa
et al. 2009). Similar spectra were also obtained by Torres et al. (2014) for the X-ray sources from the
Galactic Bulge Survey identified as low-mass X-ray binaries in their low state.

The total number of photons registered by Chandra from the error circle of AX J173548-3207 is too
small (46) to carry out detailed studies of its X-ray spectrum. It can only be said that its slope $\Gamma\simeq2.0$ is
typical of low-mass X-ray binaries and is consistent with the ASCA measurements (Sugizaki et al. 2001).

The flux measured by Chandra turns out to be a factor of $\sim3$ lower than that measured by ASCA. The same
can also be said about the absorption column density: our measurements give a value that is a factor of
$\sim4$ lower than the ASCA ($0.3$ vs $1.35$ in units $10^{22}$ cm$^{-2}$), However, the accuracy of both these
measurements is low, and they formally differ only by $2\sigma$.

\begin{figure*}
\begin{center}
\includegraphics[width=0.95\textwidth,bb=26 201 576 624,clip]{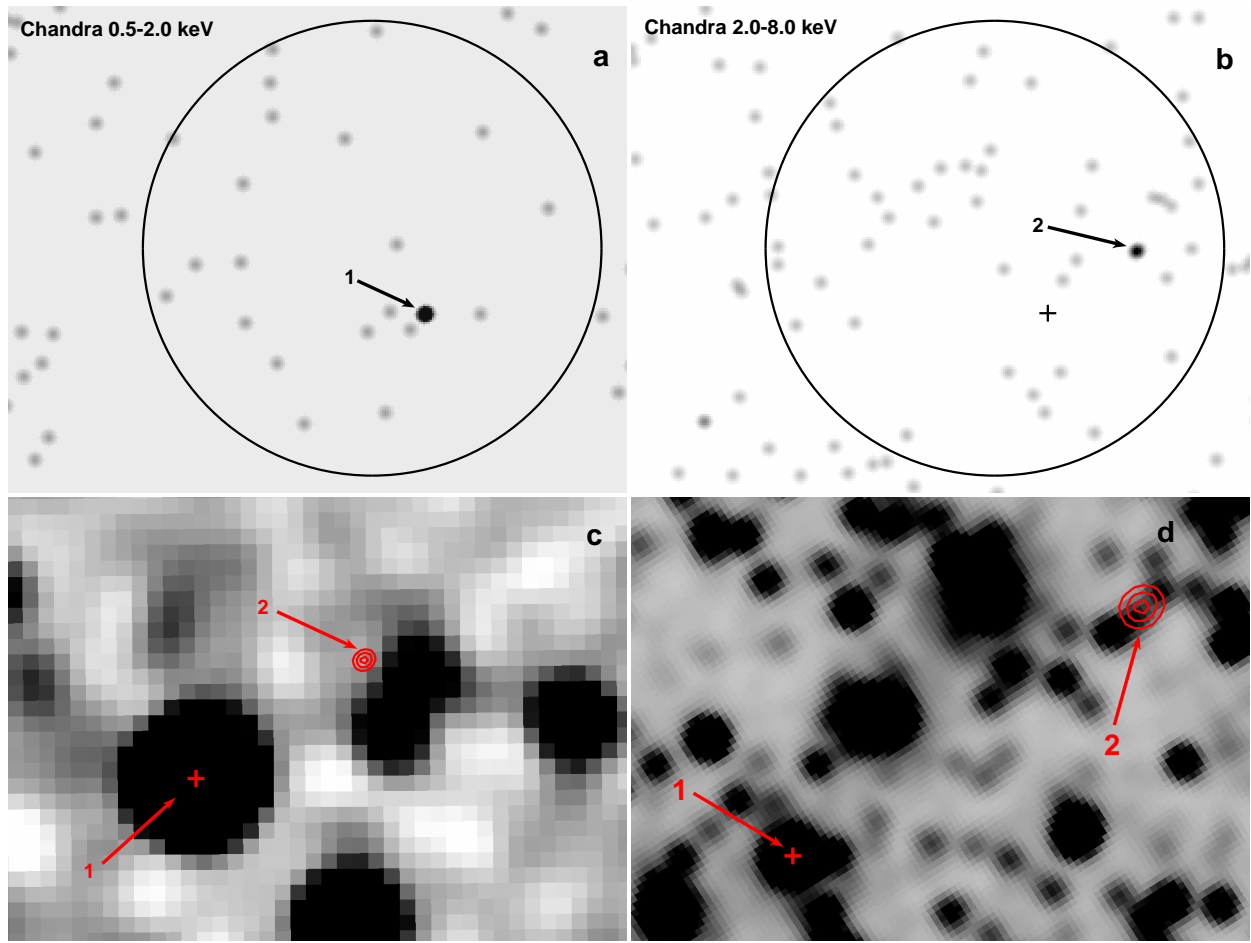} \caption{Images of the 
sky field around AX\,J173628-3141. Panels (a) and (b) show the Chandra images in the 0.5–2 and 
2–8 keV X-ray bands, respectively. The circle indicates the ASCA error circle of 
AX\,J173628-3141; the position of the soft X-ray source is indicated by the cross on panel (b). 
The lower panels show the 2MASS (c) and VVV (d) images of this field. The soft X-ray source 
coincident with the bright star 2MASS\,J17362729-3141234 is marked by the arrow and number 1. 
The error circle of the hard X-ray source is indicated by the contours and number 2.}
\label{axj173628_image}
\end{center}
\end{figure*}

The APASS\footnote{http://www.aavso.org/apass} and 2MASS (Skrutskie et al. 2006)
photometry allows one to set limits on the temperature of emitting region (most likely an accretion
disk), $T>8000$ К and the interstellar reddening in the spectrum $E(B-V)<0.4$, which corresponds to a
heliocentric distance of less than 2 kpc (Lallement et al. 2014).

\subsection{AX\,J173628-3141}

\begin{figure}
\begin{center}
\includegraphics[width=\columnwidth,bb=50 185 570 490,clip]{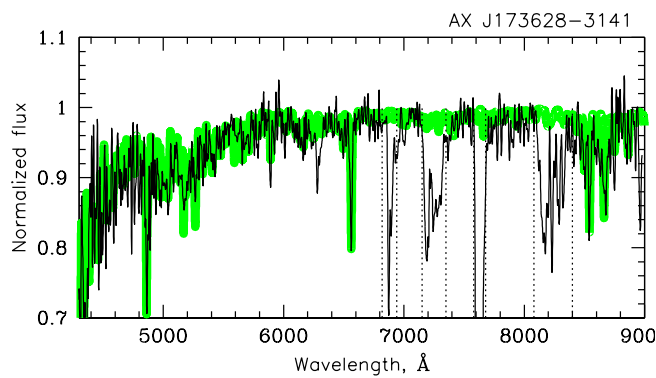}
\caption{RTT-150 optical spectrum (normalized to the continuum) of the soft X-ray counterpart to AX\,J173628-3141. For
comparison, the green line indicates the model of a red G giant with $T\sim5900$ K and $\log g\sim3.2$. The dotted straight lines
mark the absorption bands in the Earth’s atmosphere.}
\label{axj173628}
\end{center}
\end{figure}

Chandra detected two X-ray sources in the error circle of AX\,J173628-3141 (Fig.\ref{axj173628_image}; 
see also Anderson et al. 2014).
These objects differ significantly in their spectral properties, as can be clearly seen from the
comparison of the X-ray images in the 0.5–2 and 2–8 keV energy bands (Fig.\ref{axj173628_image}a,b).
One of them has a soft X-ray spectrum typical of the emission from coronally active stars (see, e.g., 
Gudel 2004). This source coincides with the bright optical star 2MASS\,J17362729-3141234 ($K\sim10.6$), 
which spectrum was taken at RTT-150. The shape of the source's spectrum allows it to be identified 
with a G giant ($\log g\sim3.2$) with a surface temperature $T\sim5900$ К (Fig.\ref{axj173628}). 
A similar estimate of the stellar temperature $T\approx6228$ K was obtained by Ammons et al. (2006) 
from the model fit to Tycho and 2MASS photometry. The object's small proper motion in the
sky, $3\pm4$ mas yr $^{-1}$ (Roeser et al. 2010), points to a large distance to it, more than several 
hundred parsecs (see, e.g., the correlations between the proper
motions and distances of objects in Warwick 2014).

\begin{figure}[htb]
\begin{center}
\includegraphics[width=0.9\columnwidth,bb=57 207 571 700,clip]{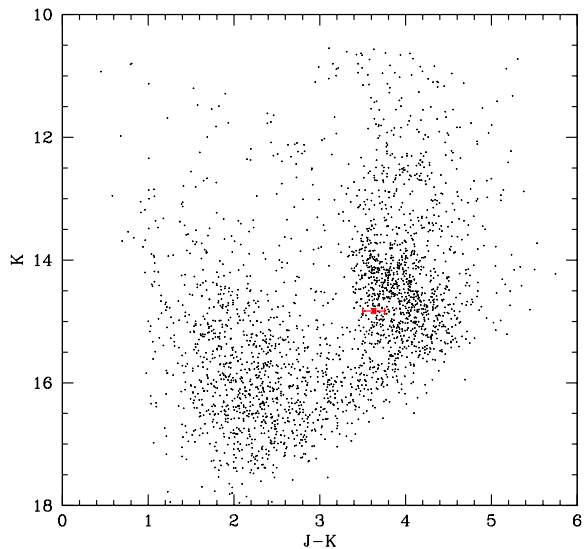}
\caption{(J-K)-K color–magnitude diagram constructed for a 2\arcmin neighborhood of the optical counterpart 
to object \#2 from the error circle of AX\,J173628-3141. The square indicates the position of the object under study.}
\label{axj173628_HK}
\end{center}
\end{figure}

We can attempt to determine the distance D to a source by the so-called spectroscopic parallax technique,
i.e., by spectroscopically estimating the stellar radius $R_\star$ and comparing the estimated luminosity
of the star with its observed flux corrected for interstellar extinction. For the source being discussed,
$R_\star\sim5.4 R_\odot$, where $R_\odot$ -- is the solar radius. Using
model stellar atmospheres from Kurucz’s library\footnote{http://www.stsci.edu/hst/observatory/crds/ k93models.html},
we can estimate the ratio $R_\star/D\sim8.5\times10^{-11}$.
Thus, the distance to the source can be estimated as $D\sim1.4$ kpc.
To refine the result obtained, it is necessary to carry out more detailed spectroscopic observations
with a better spectral resolution. At a distance of 1.4 kpc, the source’s X-ray flux 
$F_{\rm 0.5-10\,keV}\sim1.2\times10^{-14}$ \flux\ corresponds to a 
luminosity $L_{\rm 0.5-10\,keV}\sim3\times 10^{30}$ \ergs. Thus, the source is most likely 
an RS CVn object, a binary system where the mutual tidal influence of
the stars allows the convective envelope of the giant star to be continuously spun, producing an extended
corona that is an origin of X-ray emission (Walter and Bowyer 1981; Dempsey et al. 1993; Sazonov
et al. 2006).

The second X-ray source detected by Chandra in the error circle of AX\,J173628-3141, has a considerably
harder spectrum or significant absorption: all photons from it are registered by Chandra
at energies above 2 keV, while it has the largest effective area at energies below 2 keV.
The source’s spectrum can be satisfactorily (given the poor statistics, because only five photons were
recorded from the source) described by a power-law energy dependence of the photon flux density with
photoabsorption with a column density $N_{\rm H}>2\times 10^{23}$ cm$^{-2}$.

\begin{figure*}
\begin{center}
\includegraphics[width=0.95\textwidth,bb=35 307 575 486,clip]{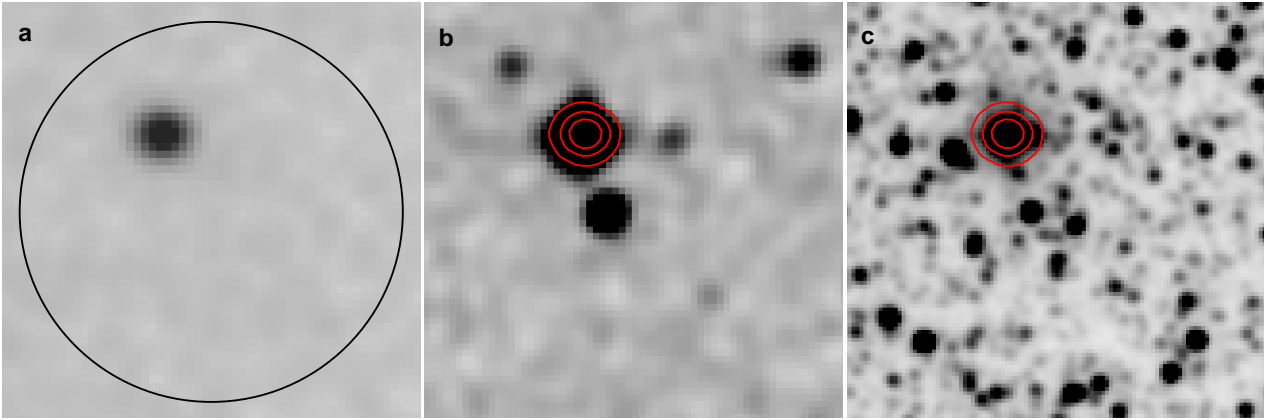}
\caption{Images of the sky field around AX\,J1739.5-2910. Panel (a) shows the Chandra image in the 0.5–8 keV X-ray band,
panel (b) shows the optical image, and panel (c) shows the 2MASS infrared image. The position of the X-ray source is marked
by the contours in the optical and infrared images.}
\label{axj17395_image}
\end{center}
\end{figure*}
\begin{figure*}
\begin{center}
\hbox{
\includegraphics[width=\columnwidth,bb=50 181 583 482,clip]{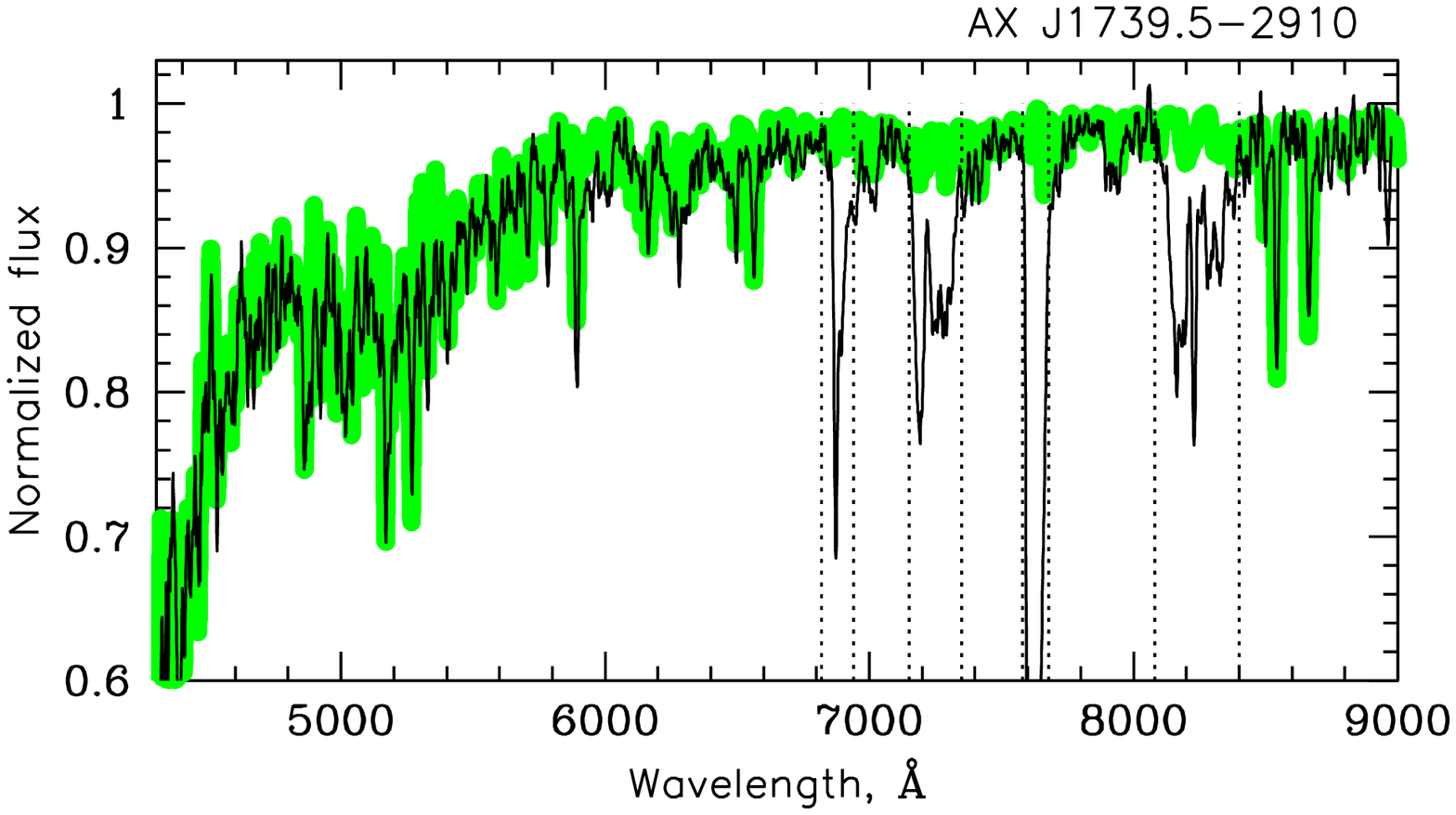}
\includegraphics[width=\columnwidth,bb=33 202 564 691,clip]{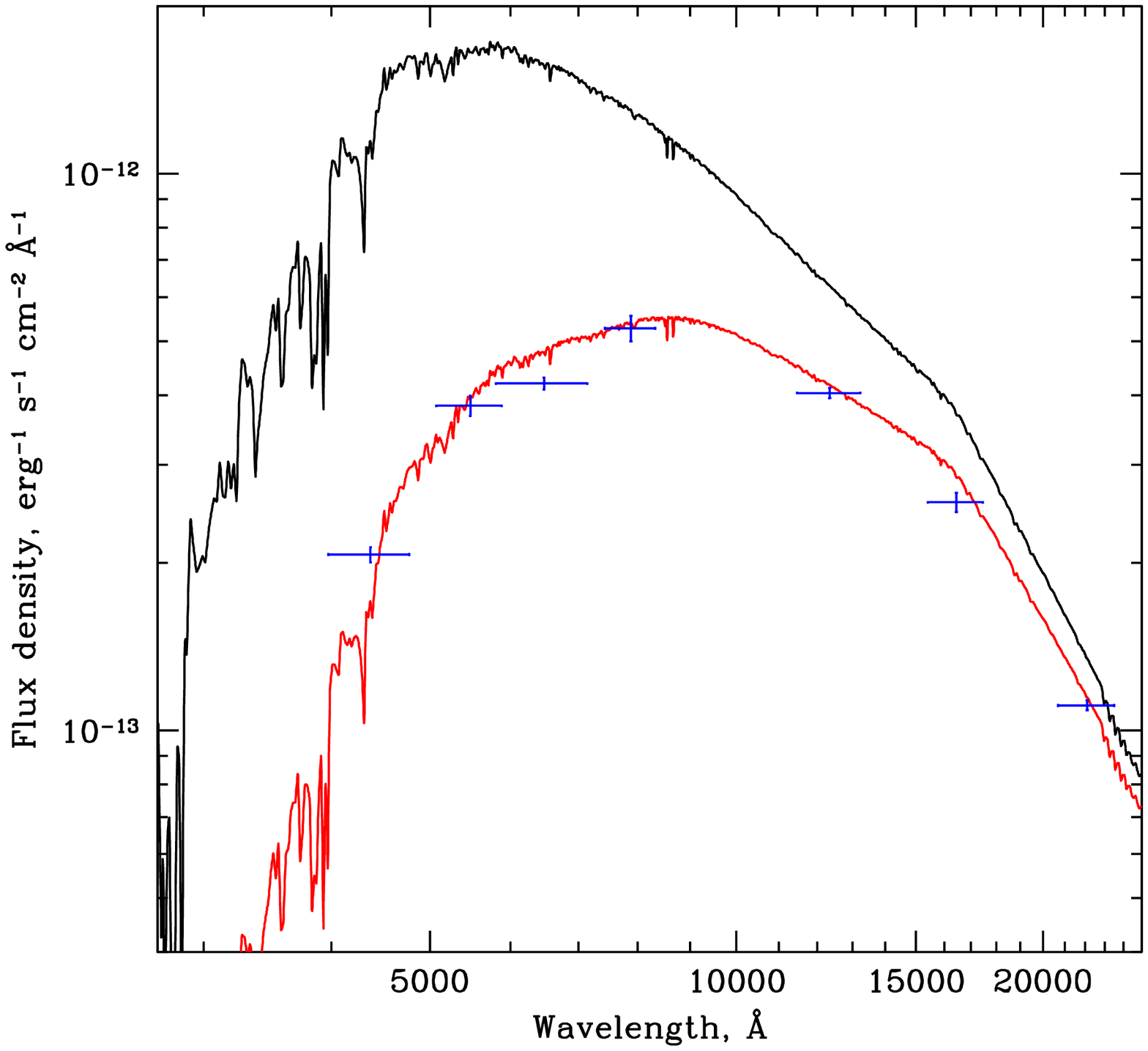}
}
\end{center}
\caption{Top: the RTT-150 optical spectrum (normalized to the continuum) of AX\,J1739.5-2910.For comparison, the
green color indicates the theoretical spectrum of a K giant  ($\log g\sim2.4$) with temperature $T=4900$ K. The
dotted straight lines mark the absorption bands in the Earth’s atmosphere. Bottom: the photometric brightness
measurements for the source in various bands (crosses) in comparison with the model of emission from a star
with temperature $T=4900$ K absorbed (red curve) and unabsorbed (black curve).}
\label{axj17395}
\end{figure*}

We investigated the properties of the second putative counterpart to AX\,J173628-3141
in the infrared using the public VVV/ESO Galactic Bulge Survey data. To achieve a higher accuracy
of determining the photometric magnitudes in such a crowded sky field, we processed the VVV observations
by the methods of PSF-photometry. For this purpose, we used the standard DAOPHOTII package
and took 2MASS as a reference catalog to obtain the photometric solution. As a result, we obtained
the object's magnitudes in three bands: $m_{\rm Ks}=14.83\pm0.38$, $m_{\rm H}=15.85\pm0.04$, $m_{\rm J}=18.46\pm0.12$.
It can be seen from the color–magnitude diagram (Fig.\ref{axj173628_HK}), constructed for a 2\arcmin\
neighborhood of the studied object that it is close to the positions of red
giants in the Galactic center region and, thus, may be at a distance of $\sim 8500-9000$ pc.
Taking into account a very significant absorption in the source's
X-ray spectrum, it can be suggested that the source is a symbiotic binary: a white dwarf accreting matter
from the stellar wind of its companion, a red giant.

Deep Chandra or XMM-Newton X-ray observations with the goal of accurately measuring the
spectrum of this object and primarily the absorption column density $N_{\rm H}$ are needed to ultimately establish
its nature.

\subsection{AX\,J1739.5-2910}

The sky field around the X-ray source АХ\,J1739.5-2910 was observed by Chandra (Evans et al. 2010),
which allowed its astrometric position to be improved (Fig. \ref{axj17395_image}).
There is only one optical (or infrared) object -- 2MASS\,J17393122-2909532
in the error circle of the source detected by Chandra. To determine the
nature of this object, we analyzed the set of its photometric measurements in the APASS surveys
(Henden et al. 2012) and took its optical spectrum at RTT-150 (Fig.\ref{axj17395}).
The spectral features visible in this spectrum allow the type of the star to be
reliably determined, a G–K giant with $T\sim4900$ K and $\log g\sim2.4$ (see also Torres et al. 2006). Our
photometry for the object points to a certain reddening (due to interstellar extinction), 
$E(B-V)\sim0.5$ (Fig.\ref{axj17395}) in the spectrum.

It can be concluded from the data obtained that we most likely deal with a giant К1–2 III star at a
distance of $\sim 400$ pc (the stellar radius estimate is $R_\star\sim12 R_\odot$  and the ratio
$R_\star/D\sim6.7\times10^{-10}$). At such a distance, the necessary reddening $E(B-V)\sim0.5$ can already 
be accumulated toward the Galactic Center (see, e.g., Lallement et al. 2014). It should be noted that 
similar estimates of the spectral type and distance to this optical star were obtained by Pickles and 
Depagne (2010) without any relation to X-ray observations.

\begin{figure*}
\begin{center}
\includegraphics[width=0.95\textwidth,bb=26 201 576 650,clip]{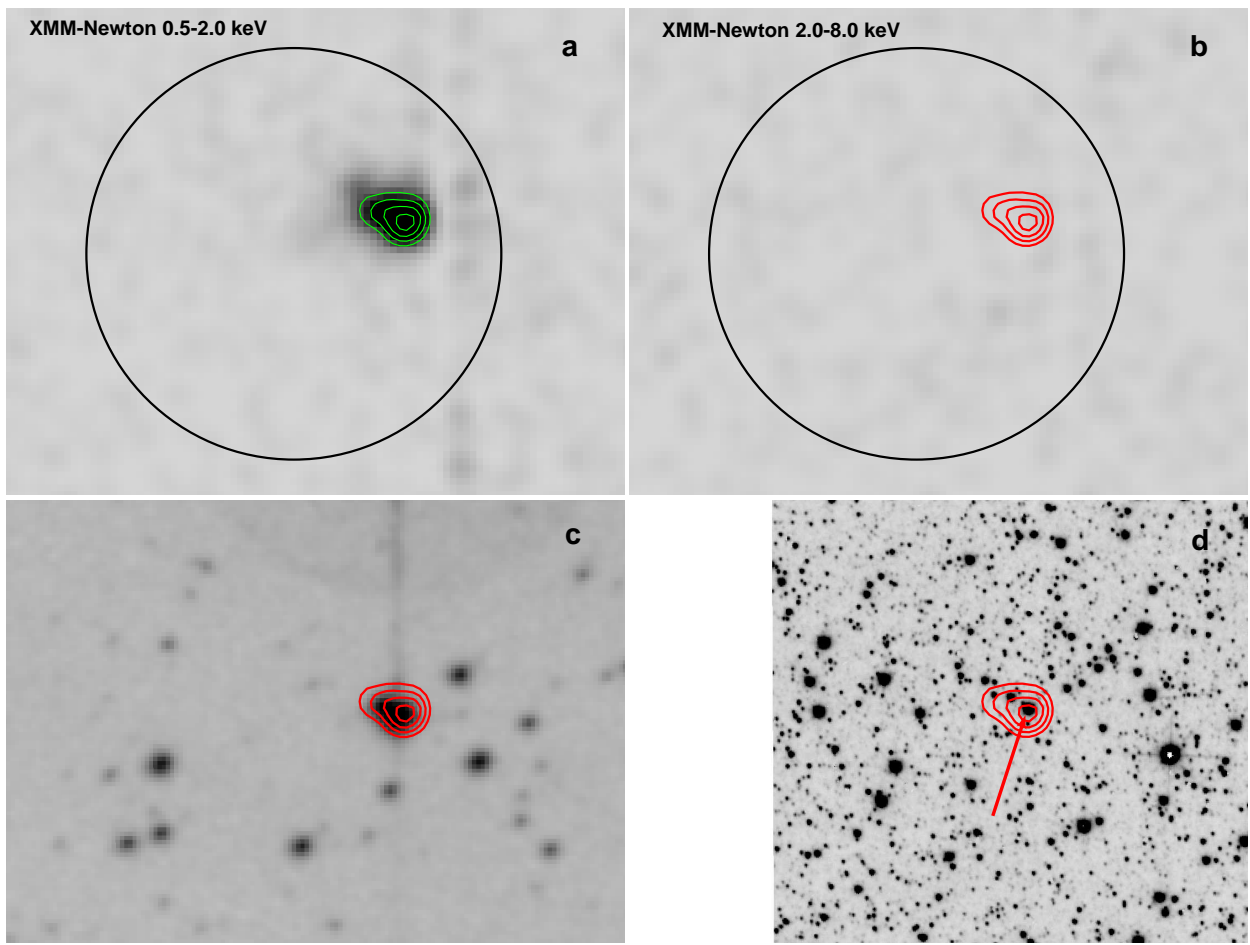}
\end{center}
\caption{Images of the sky field around AX\,J1740.4-2856. Panels (a) and (b)
show the XMM-Newton images in the 0.5–2 and 2–8 keV energy bands, respectively.
The circle indicates the ASCA localization accuracy of AX\,J1740.4-2856. The lower
panels show the 2MASS (c) and VVV (d) images of this field. The contours correspond
to the X-ray intensity distribution in the 0.5–2 keV
energy band. The arrow indicates the infrared companion.}
\label{axj17404_image}
\end{figure*}
\begin{figure*}
\begin{center}
\includegraphics[width=1.05\columnwidth,bb=50 175 590 490,clip]{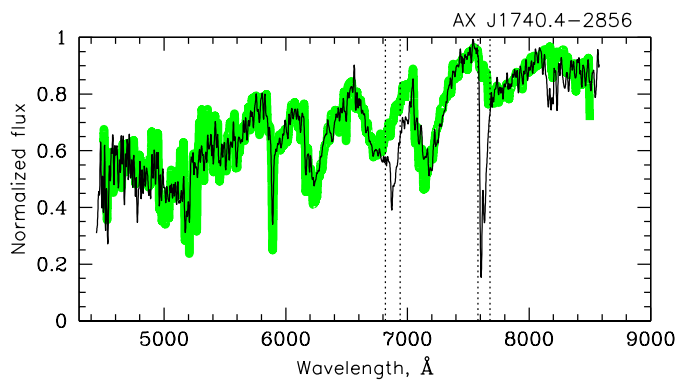}
\caption{RTT-150 optical spectrum (normalized to the continuum) of AX\,J1740.4-2856.For comparison, the green color
indicates the model of a red M-dwarf with $T\sim3800$ K, $\log g\sim2.5$. The dotted straight linesmark the absorption bands
in the Earth’s atmosphere.}
\label{axj17404}
\end{center}
\end{figure*}

The source was observed several times by Chandra
and XMM-Newton in the X-ray energy band, which allowed its spectrum and X-ray flux to be measured
at different times. It can be seen from the table that the source flux is virtually
the same in all energy bands, as is the shape of its spectrum. Moreover, the results of our measurements
agree well with those of the ASCA ones.
The registered flux $F_{\rm 0.5-10 KeV}\sim10^{-12}$ \flux\ at a distance $\sim400$ pc
corresponds to a luminosity $L_{\rm 0.5-10 keV}\sim1.2\times10^{31}$ \ergs .
Note that AX\,J1739.5-2910 is one of the brightest objects detected during the
Chandra Galactic Bulge Survey (the source was designated as CX4=CXOGBS\,J173931.2–290952 in
this survey; Hynes et al. 2012). In addition, an optical variability of this source with a period of about 16 days
was observed (Pojmanski 2002; Hynes et al. 2012), suggesting that this object belongs to the class of Xray-
active RS CVn binaries.

\subsection{AX\,J1740.4-2856}

\begin{figure*}
\begin{center}
\includegraphics[width=0.95\textwidth,bb=35 307 575 486,clip]{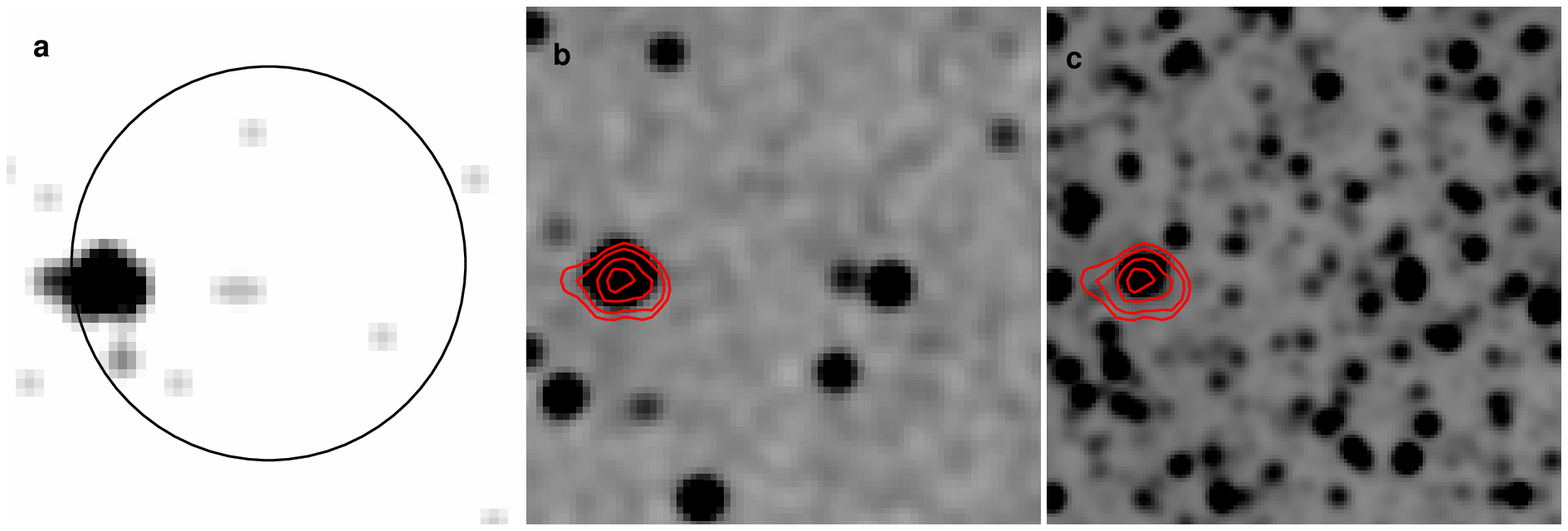}
\end{center}
\caption{Images of the sky field around AX\,J1740.5-2937. Panel (a) shows the {\em XRT}/Swift image in the 0.5–8 keV energy
band, panel (b) shows the optical image, and panel (c) shows the infrared (2MASS) image. The position of the X-ray source
is marked by the contours in the optical and infrared images.}
\label{axj17405_image}
\end{figure*}

AX\,J1740.4-2856, detected by ASCA can be unambiguously identified with the X-ray source detected
by ХММ-Newton in its error circle, 3XMM J174023.9-285652 (from the 3XMM-DR4 survey).
This position allows it to be associated with the optical/infrared object 2MASS J17402384-2856527  
(Fig.\ref{axj17404_image}). It is important to note that the source is
detected with confidence in the 0.5–2 keV energy
band and is absent in harder X-rays (Figs.\ref{axj17404_image}a,b), directly suggesting that its spectrum is
soft. Indeed, the latter can be described by the model of a multi-temperature plasma with a maximum
temperature $kT\simeq0.9$ KeV.
Such a spectral shape is typical of the emission from coronally active stars.
Note that the X-ray flux measured during two XMM-Newton observations is a factor of $\sim3$ lower than that
detected previously by ASCA.

The optical spectrum of this source obtained at RTT-150 is shown in Fig.\ref{axj17404}.It can be described by the
model of emission from an early-M star with $T\sim3800$ K with a poorly determined luminosity class, $\log g\sim1-4$.

The APASS and 2MASS photometric brightness
measurements for the object give an estimate of the
ratio of the star’s radius to its distance $R_\star/D\sim2.4\times10^{-10}$,and an interstellar reddening $E(B-V)<0.3$.
The distance for a giant star would be more than 3–4 kpc toward the Galactic center, which would
inevitably give rise to a significant reddening in the spectrum (Lallement et al. 2014); therefore, the star
is most likely an ordinary M dwarf. In that case, assuming the typical radii of M0–2 stars to be $R_\star\sim0.6R_\odot$, the distance to the star will be $\sim56$ pc. For such a distance, the observed X-ray flux $F_{\rm 0.5-10
KeV}\sim3\times10^{-13}$ \flux\  corresponds to a luminosity $L_{\rm 0.5-10 keV}\sim 10^{29}$ \ergs.
The ratio of the source’s X-ray and infrared fluxes $F_{\rm x}/F_{\rm j}\sim 10^{-3}$ is typical for
objects of this class (Agueros et al. 2009).

Interestingly, according to the Optical Monitoring Camera (OMC) observations onboard INTEGRAL,
this source exhibited bright flares up to $m_{\rm V}\sim$9 in $V$; at the same time, the normal brightness of this source
is $m_{\rm V}\approx13$ (Alfonso-Garz{\'o}n et al. 2012). Flares of such a brightness are occasionally observed in
M dwarfs (Kowalski et al. 2009).

\subsection{AX\,J1740.5-2937}

\begin{figure}
\begin{center}
\includegraphics[width=\columnwidth,bb=50 175 570 490,clip]{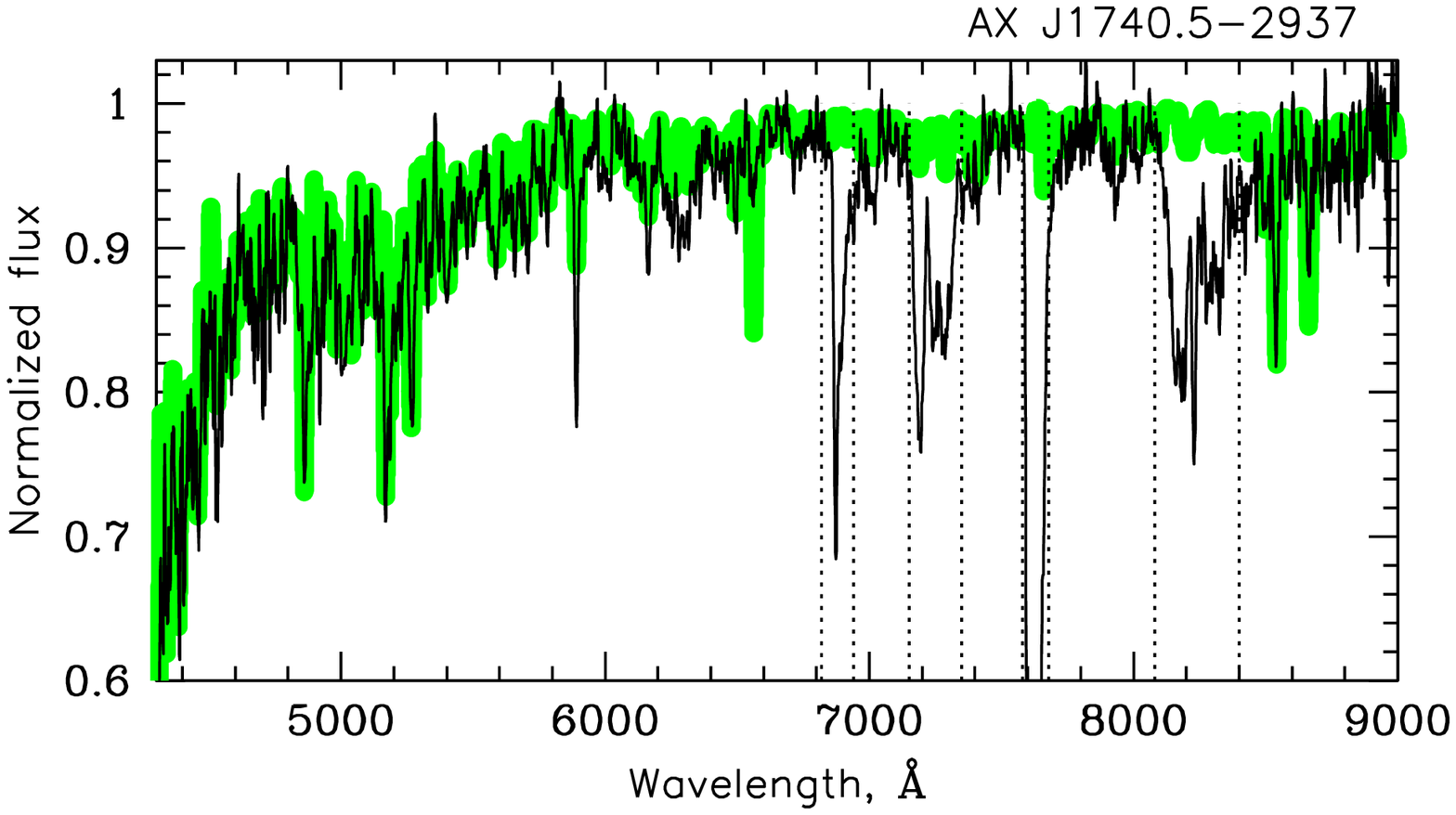}
\caption{RTT-150 optical spectrum (normalized to the continuum) of AX\,J1740.5-2937. For comparison, the green color
indicates the theoretical spectrum of a G–K subgiant  ($\log g\sim3.1$) with temperature $T=5400$ K.The dotted straight lines mark the absorption bands in the Earth’s atmosphere.}
\label{axj17405}
\end{center}
\end{figure}

The optical counterpart to AX\,J1740.5-2937 (2MASS\,J17403458-2937438) has been determined
owing the detection of only one X-ray object within its error circle (Fig. \ref{axj17405_image}), for which an accurate astrometric
position was obtained (Evans et al. 2014). Its optical spectrum measured at RTT-150 (Fig. \ref{axj17405}), is
satisfactorily described by the model of a G subgiant ($\log g\sim3.1$) with a temperature $T\sim5400$ К.

\begin{figure*}
\begin{center}
\includegraphics[width=0.95\textwidth,bb=35 307 575 486,clip]{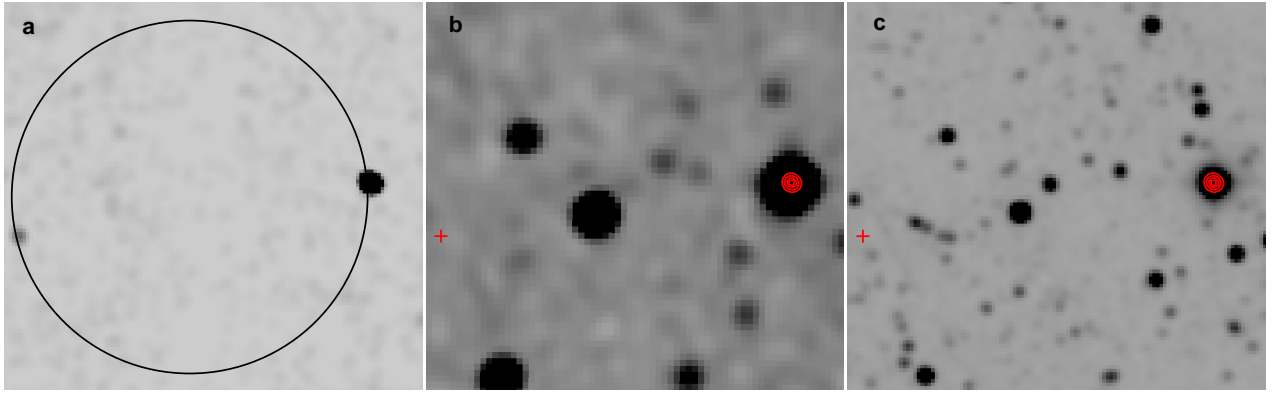}
\end{center}
\caption{Images of the sky field around AX\,J1743.9-2846. Panel (a) shows the Chandra image in the 0.5–8 keV energy band,
panel (b) shows the optical image, and panel (c) shows the infrared (2MASS) image. The positions of the bright and faint
X-ray sources are marked by the contours and cross, respectively, in the optical and infrared images.}
\label{axj17439_image}
\end{figure*}
The fit to the photometric measurements from Pickles and Depagne (2010) gives an estimate of the
spectral type of the star as K2V at a distance of 91 pc. To all appearances, such a result stems from
the fact that these authors disregarded the interstellar reddening, which led to an underestimate of the stellar
temperature. Our fit to the photometric measurements from APASS and 2MASS by the model of
a star with a temperature of 5400 K allows us to estimate the interstellar reddening, $E(B-V)\sim0.2$
(consistent with the estimates of the column density
from the interstellar photoabsorption in X-ray data; see the table), and the ratio of the distance to the
star to its radius, $R_\star/D\sim2.0\times10^{-10}$.  At the star’s radius $R_\star\sim6R_\odot$ the 
distance to it turns out to be $D\sim670$ pc.

For such a distance, the source’s X-ray flux $F_{\rm 0.5-10\,KeV}\simeq6\times10^{-13}$ \flux\ corresponds
to a luminosity $L_{\rm 0.5-10 keV}\sim2-5\times10^{31}$ \ergs\ in the 0.5–10 keV energy band. Note that, according
to the ASCA data, the flux from the source was approximately the same ($8\times10^{-13}$ \flux\ in the 0.7–10 keV
energy band), while the spectrum
had a slope of 4.6. In our work, we used the more physical model of a multi-temperature plasma with a
maximum temperature $kT\sim0.75$ keV (see the table) to fit the spectrum.

\subsection{AX\,J1743.9-2846}

AX\,J1743.9-2846 fell several times within the Chandra and ХММ-Newton fields of view, which allowed
its astrometric position to be improved (Evans et al. 2010). Two relatively bright sources were
detected within a circle 50\arcsec\ in radius, with both being virtually at its edge (Fig.\ref{axj17439_image}).
One of these objects (CXO\,J174351.3-284637 = 3ХММ\,J174351.2-284637 = 1SXPS\,J174351.2-284636)  is brighter than the
other approximately by a factor of 10, suggesting that it corresponds to  AX\,J1743.9-2846.
The source CXO\,J174351.3-284637, in turn, coincides with the bright star 2MASS\,J17435129-2846380, whose
photometry makes it possible to estimate its spectral type, К3 III, and distance, $\sim1.1$ kpc (Pickles and Depagne 2010).

\begin{figure}
\begin{center}
\includegraphics[width=\columnwidth,bb=50 175 570 490,clip]{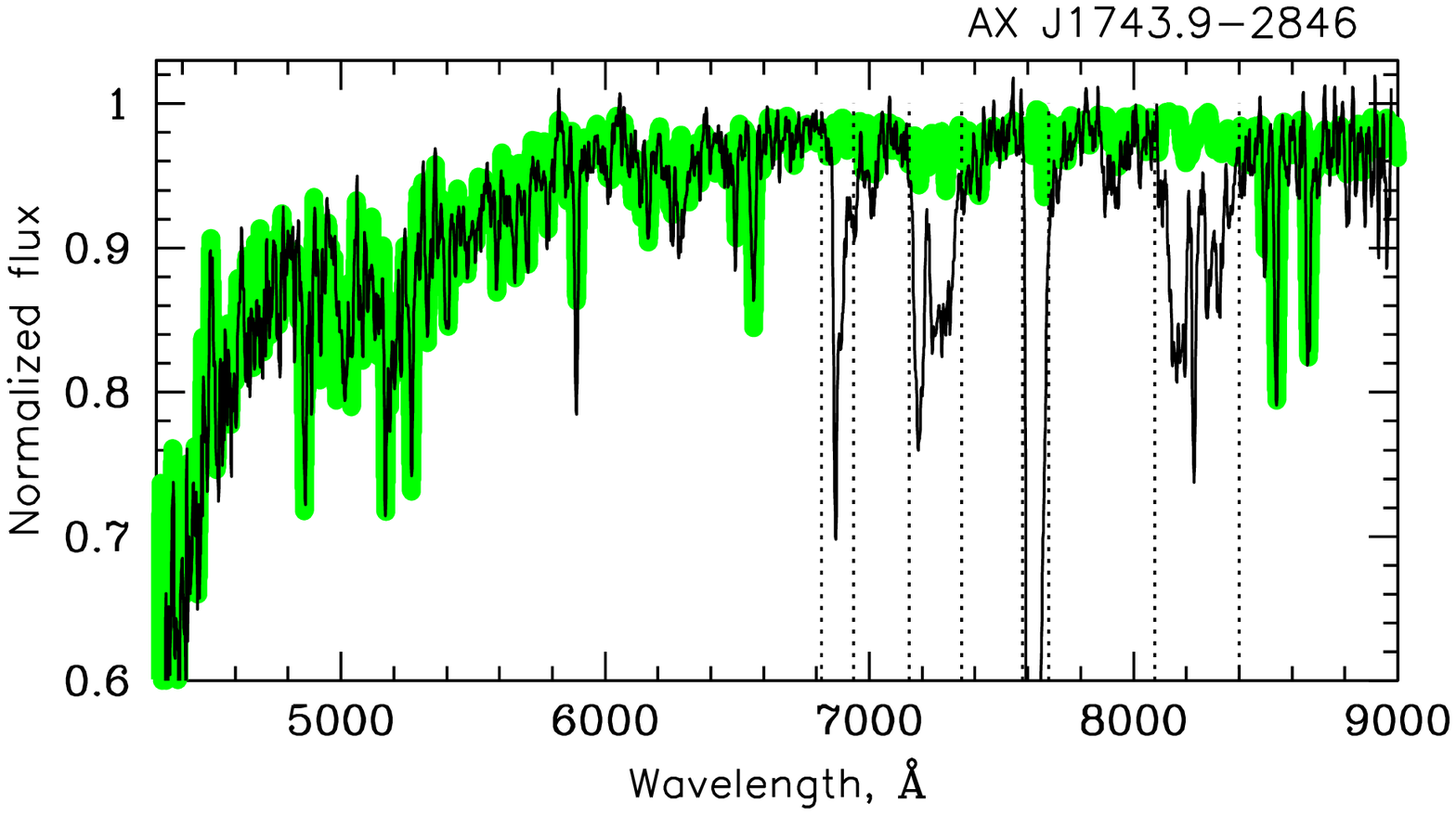}
\caption{RTT-150 optical spectrum (normalized to the continuum) of AX\,J1743.9-2846. For comparison,
the green color indicates the theoretical spectrum of a G subgiant ($\log g\sim2.9$) with
temperature $T=5350$ K. The dotted straight lines mark the absorption bands in the Earth’s
atmosphere. }
\label{axj17439}
\end{center}
\end{figure}

Our RTT-150 spectroscopy (Fig.\ref{axj17439}), points to a
slightly higher stellar temperature, $\sim$5350 K, and a
surface gravity $\log g\sim2.9$. As with the preceding
source AX\,J1740.5-2937, this may be because the
stellar temperature estimate obtained from the photometry
of bright stars disregards the possible contribution
of interstellar reddening. In our case, we can
estimate the interstellar reddening as $E(B-V)\sim0.5$ and the ratio
$R_\star/D\sim3.3\times10^{-10}$. For the star’s estimated radius $R_\star\sim7R_\odot$ the distance to it will be $D\sim 480$ pc.

The source’s X-ray spectrum can be well described
by a power law with a break modified by
absorption at low energies (see Fig. \ref{spectra} and the table). During the ASCA observations,
AX\,J1743.9-2846 exhibited a significant flux variability (from  $8\times10^{-13}$ to $10^{-11}$ \flux\
in the 0.7-10 keV energy band). The Chandra and XMM-Newton observations show that the flux from the source and its
variability amplitude were considerably lower than those during the ASCA observations (see the table).
At the same time, the shape of the spectrum and its parameters determined from Chandra and XMM-Newton
data are similar (except for the observation 0112971201, where the parameters turned out to be
slightly different, but the significance of this difference is low because of insufficient statistics).

The XMM-Newton and Chandra X-ray fluxes correspond to a luminosity $L_{\rm 0.5-10\,KeV}\sim(3-9)\times10^{31}$ \ergs,
for a distance to the source of 1.1 kpc
and $L_{\rm 0.5-10\,keV}\sim(0.7-2)\times10^{31}$ \ergs\ for a distance of
0.5 kpc. For RS CVn stars, the X-ray luminosity we derived is at the brightest edge of the distribution; in
particular, it can be higher than the X-ray luminosity of known RS CVn binaries within 50 pc of the Sun
(see, e.g., Makarov 2003).

\bigskip

\section{CONCLUSIONS}

In this paper, we analyzed the X-ray and optical observations of six sources from the ASCA Galactic
Center and Galactic plane surveys: AX\,J173548-3207, AX\,J173628-3141, AX\,J1739.5-2910,
AX\,J1740.4-2856, AX\,J1740.5-2937, and AX\,J1743.9-2846. These surveys are presently the deepest largearea
(about 45 square degrees) surveys performed by focusing X-ray telescopes at energies above
2 keV, the energy band that is weakly affected by interstellar photoabsorption. We obtained an unambiguous
identification for five of the six objects; accurate Chandra X-ray measurements for one more
object (AX\,J173628-3141) reveal two X-ray sources within its localization region (soft and hard) of comparable
brightness in the 0.5–8 keV energy band. Three of the five sources with an unambiguous identification
are active G–K stars, presumably RS CVn binaries. One of the two objects in the error circle of
AX\,J173628-3141 is a coronally active G star, while the other object is a symbiotic star, a red giant with an
accreting white dwarf, presumably located in the Galactic center region. AX\,J1740.4-2856 is a coronally active red
M0-2 dwarf; one more source, AX\,J173548-3207, is presumably an X-ray binary in our Galaxy (most
likely a low-mass one) in its low state. We estimated the distances and corresponding luminosities in the
soft X-ray band (0.5–10 keV) for all sources, with none of them having been detected at a statistically
significant level at energies >20 keV based on INTEGRAL
data.

\bigskip 

This work was supported by the Russian Science Foundation (grant 14-22-00271). We used the data retrieved
from the archives of the Goddard Space Flight Center (NASA) and the European Southern Observatory.
We thank TUBITAK (Turkey), the
Space Research Institute of the Russian Academy of Sciences, and the Kazan Federal University for their
support in using the Russian–Turkish 1.5-m telescope (RTT-150). We are grateful to E.M. Churazov,
who developed the IBIS/INTEGRAL data analysis methods and provided the software. V.V. Shimansky
and I.F. Bikmaev thank the Russian Foundation for Basic Research (project no. 13-02-00351) for partial
support.

\vspace{1cm}

\centerline{REFERENCES}

\bigskip

\parindent=0mm

Ag{\"u}eros M.~A., Anderson S.~F., Covey K., Hawley S., Margon B., Newsom E., Posselt B., Silvestri N., 
et al., \apjs\ {\bf 181}, 444, (2009)

Alfonso-Garz{\'o}n J., Domingo A., Mas-Hesse J.~M., Gim{\'e}nez A., \aap\ {\bf 548}, A79, (2012)

Ammons S.~M., Robinson S.~E., Strader J., Laughlin G., Fischer D., Wolf A., \apj\ {\bf 638}, 1004, (2006)

Anderson G.~E., Gaensler B., Kaplan D., Slane P., Muno M., Posselt B., Hong J., Murray S., et al., \apjs\ {\bf 212}, 13, (2014)

Bassa C.~G., Jonker P.~G., Steeghs D., Torres M.~A.~P., \mnras\ {\bf 399}, 2055, (2009)

Baumgartner W., Tueller J., Markwardt C., Skinner G., Barthelmy S., Mushotzky R., Evans P., Gehrels N., \apjs\ {\bf 207}, 19, (2013)

Bodaghee A., Tomsick J.~A., Rodriguez J., James J.~B., \apj\ {\bf 744}, 108, (2012)

Churazov E., Sunyaev R., Isern J., Kn{\"o}dlseder J., Jean P., Lebrun F., Chugai N., Grebenev S., et al., Nature {\bf 512}, 406, (2014)

Degenaar N., Starling R., Evans P., Beardmore A., Burrows D., Cackett E., Campana S., Grupe D., et al., \aap\ {\bf 540}, A22, (2012)

Dempsey R.~C., Linsky J.~L., Fleming T.~A., Schmitt J.~H.~M.~M., \apjs\ {\bf 86}, 599, (1993)

Ebisawa K., Tsujimoto M., Paizis A., Hamaguchi K., Bamba A., Cutri R., Kaneda H., Maeda Y., et al., \apj\ {\bf 635}, 214, (2005)

Evans I.~N., Primini F.~A., Glotfelty K.~J., Anderson C., Bonaventura N., Chen J.~C., Davis J.~E.; Doe S., 
et al., \apjs\ {\bf 189}, 37 (2010)

Evans P.~A., Osborne J., Beardmore A., Page K., Willingale R., Mountford C., Pagani C., Burrows D., 
et al., \apjs\  {\bf 210}, 8, (2014)

Grimm H.-J., Gilfanov M., Sunyaev R., \aap\ {\bf 391}, 923, (2002)

G{\"u}del M., A\&ARv {\bf 12}, 71, (2004)

Hands A.~D.~P., Warwick R.~S., Watson M.~G., Helfand D.~J., \mnras\ {\bf 351}, 31, (2004)

Henden A.~A., Levine S.~E., Terrell D., Smith T.~C., Welch D., JAVSO {\bf 40}, 430, (2012)

Hynes R., Wright N., Maccarone T., Jonker P., Greiss S., Steeghs D., Torres M.~A.~P., Britt C., et al., \apj\ {\bf 761}, 162, (2012)

Kim D.-W., Cameron R., Drake J., Evans N.~R., Freeman P., Gaetz T., Ghosh H., Green P., et al., \apjs\ {\bf 150}, 19, (2004)

Kowalski A.~F., Hawley S.~L., Hilton E.~J., Becker A.~C., West A.~A., Bochanski J.~J., Sesar B., \aj\ {\bf 138}, 633, (2009)

Krivonos R., Revnivtsev M., Lutovinov A., Sazonov S., Churazov E., Sunyaev R., \aap\ {\bf 475}, 775, (2007)

Krivonos R., Revnivtsev M., Tsygankov S., Grebenev S., Churazov E., Sunyaev R., \aap\ {\bf 519}, A107 (2010)

Krivonos R., Tsygankov S., Lutovinov A., Revnivtsev M., Churazov E., Sunyaev R., \aap\ {\bf 545}, A27, (2012)

Lallement R., Vergely J.-L., Valette B., Puspitarini L., Eyer L., Casagrande L., \aap\ {\bf 561}, A91, (2014)

Lutovinov A., Revnivtsev M., Gilfanov M., Shtykovskiy P., Molkov S., Sunyaev R., \aap\ {\bf 444}, 821, (2005)

Lutovinov A.~A., Revnivtsev M.~G., Tsygankov S.~S., Krivonos R.~A., \mnras\  {\bf 431}, 327, (2013)

Makarov V.~V., \aj\ {\bf 126}, 1996, (2003)

Pickles A., Depagne {\'E}., Publ. Astron. Soc. Japan {\bf 122}, 1437, (2010)

Pojmanski G., AcA {\bf 52}, 397, (2002)

Revnivtsev M., Sazonov S., Jahoda K., Gilfanov M., \aap {\bf 418}, 927, (2004)

Revnivtsev M., Lutovinov A., Churazov E., Sazonov S., Gilfanov M., Grebenev S., Sunyaev R., \aap\ {\bf 491}, 209, (2008)

Revnivtsev M., Sazonov S., Churazov E., Forman W., Vikhlinin A., Sunyaev R., Nature {\bf 458}, 1142 (2009)

Roeser S., Demleitner M., Schilbach E., \aj\ {\bf 139}, 2440, (2010)

Sazonov S., Revnivtsev M., Gilfanov M., Churazov E., Sunyaev R., \aap\ {\bf 450}, 117, (2006)

Sakano M., Koyama K., Murakami H., Maeda Y., Yamauchi S., \apjs\ {\bf 138}, 19, (2002)

Saxton R.~D., Read A.~M., Esquej P., Freyberg M.~J., Altieri B., Bermejo D., \aap\ {\bf 480}, 611, (2008)

Skrutskie M., Cutri R., Stiening R., Weinberg M., Schneider~S., Carpenter J., Beichman C., Capps R., et al., 
\aj\ {\bf 131}, 1163, (2006)

Sugizaki M., Mitsuda K., Kaneda H., Matsuzaki K., Yamauchi S., Koyama K., \apjs\ {\bf 134}, 77, (2001)

Torres C.~A.~O., Quast G.~R., da Silva L., de La Reza R., Melo C.~H.~F., Sterzik M., \aap\ {\bf 460}, 695, (2006)

Torres M.~A.~P., Jonker P., Britt C., Johnson C., Hynes R., Greiss S., Steeghs D., Maccarone T., et al., \mnras\ {\bf 440}, 365, (2014)

Voges W., Aschenbach B., Boller Th., Br{\"u}ninger H., Briel U., Burkert W., Dennerl K., Englhauser J., 
et al., \aap\ {\bf 349}, 389, (1999)

van den Berg M., Hong J.~S., Grindlay J.~E., \apj\ {\bf 700}, 1702, (2009)

Wang Q.~D., Gotthelf E.~V., Lang C.~C., Nature {\bf 415}, 148, (2002)

Walter F.~M., Bowyer S., \apj\ {\bf 245}, 671, (1981)

Warwick R.~S., \mnras\ {\bf 445}, 66, (2014)

Winkler C., Courvoisier T.J.-L., Di Cocco G., Gehrels, N., Gimenez, A., Grebenev, S., Hermsen, W., 
Mas-Hesse, J. M., Lebrun, F., et al., Astron. Astrophys. 411, L1 (2003)

\end{document}